\RequirePackage{fix-cm}
\documentclass[smallextended]{svjour3}       
\smartqed  

\usepackage{graphicx}
\usepackage{amssymb}
\usepackage{amsmath}
\usepackage{appendix}
\usepackage{array,textcomp}
\usepackage[latin1]{inputenc}
\usepackage{tikz}
\usetikzlibrary{backgrounds,calc}

\pgfkeys{/tikz/.cd,
  timespan/.store in=\timespan,
  timespan=Week,
  timeline width/.store in=\timelinewidth,
  timeline width=10,
  timeline height/.store in=\timelineheight,
  timeline height=1,
  timeline offset/.store in=\timelineoffset,
  timeline offset=0.05,
  initial week/.store in=\initialweek,
  initial week=1,
  end week/.store in=\endweek,
  end week=2,
  time point/.store in=\timepoint,
  time point=0.5,
  between day/.style args={#1 and #2 in #3}{
    initial week=#1,
    end week=#2,
    time point=#3,
  },
  between week/.style args={#1 and #2 in #3}{
    initial week=#1,
    end week=#2,
    time point=#3,
  },
  between month/.style args={#1 and #2 in #3}{
    initial week=#1,
    end week=#2,
    time point=#3,
  },
  between year/.style args={#1 and #2 in #3}{
    initial week=#1,
    end week=#2,
    time point=#3,
  },
  involvement degree/.store in=\involvdegree,
  involvement degree=2cm,
  phase color/.store in=\phasecol,
  phase color=red!50!orange,
  phase appearance/.style={
    circle,
    opacity=0.3,
    minimum size=\involvdegree,
    fill=\phasecol
  },
}
\newif\ifcustominterval
\pgfkeys{/tikz/timeline/.cd,
  custom interval/.is if=custominterval,
  custom interval=false,
}

\pgfkeys{/tikz/milestone/.cd,
  at/.store in=\msstartpoint,
  at=phase-1.north,
  circle radius/.store in=\milestonecircleradius,
  circle radius=0.1cm,
  direction/.store in=\msdirection,
  direction=90:2cm,
  text/.store in=\mstext,
  text={},
  text options/.code={\tikzset{#1}},
}

\newcommand{\timeline}[2][]{
  \pgfkeys{/tikz/timeline/.cd,#1}
  \draw[fill,opacity=0.8] (0,0) rectangle (\timelinewidth,\timelineheight);
  \shade[top color=black, bottom color=white,middle color=black!20]
    (0,0) rectangle (\timelinewidth,-\timelineoffset);
  \shade[top color=white, bottom color=black,middle color=black!20]  
    (0,\timelineheight) rectangle (\timelinewidth,\timelineheight+\timelineoffset);

  \ifcustominterval%
    \foreach \smitem [count=\xi] in {#2}  {\global\let\maxsmitem\xi}%
  \else%
    \foreach \smitem [count=\xi] in {1,...,#2}  {\global\let\maxsmitem\xi}%
  \fi%
  
  \pgfmathsetmacro\position{\timelinewidth/(\maxsmitem+1)}
  \node at (0,0.5\timelineheight)(\timespan-0){\phantom{Week 0}};
 
  \ifcustominterval%
    \foreach \x[count=\xi] in {#2}{%
      \node[text=white,text depth=0pt]at +(\xi*\position,0.5\timelineheight) (\timespan-\xi) {\timespan\ \x};%
    }%
  \else%
    \foreach \x[count=\xi] in {1,...,#2}{%
      \node[text=white, text depth=0pt]at +(\xi*\position,0.5\timelineheight) (\timespan-\xi) {\timespan\ \x};%
    }%
  \fi%
}

\newcounter{involv}
\newcommand{\phase}[1]{
\stepcounter{involv}
\node[phase appearance,#1] 
 (phase-\theinvolv)
 at ($(\timespan-\initialweek)!\timepoint!(\timespan-\endweek)$){};
}

\newcommand{\initialphase}[1]{
\node[phase appearance,#1,anchor=west,between week=0 and 1 in 0,] 
 (phase-\theinvolv)
 at ($(\timespan-0)!0!(\timespan-1)$){};
\setcounter{involv}{0} 
}

\newenvironment{phases}{\begin{pgfonlayer}{background}}{\end{pgfonlayer}}

\newcommand{\addmilestone}[1]{
\pgfkeys{/tikz/milestone/.cd,#1}
\draw[double,fill] (\msstartpoint) circle [radius=\milestonecircleradius];
\draw(\msstartpoint)--++(\msdirection)node[/tikz/milestone/text options]{\mstext};
}

\usepackage{verbatim}
\usetikzlibrary{bayesnet}
\usetikzlibrary{arrows}
\usetikzlibrary{positioning,calc,arrows.meta}
\usepackage{color}
\usepackage{caption}
\usetikzlibrary{fit,positioning,arrows,automata,shapes,arrows,decorations.markings}
\definecolor{richmaroon}{rgb}{0.89, 0.15, 0.42}
\definecolor{antiquefuchsia}{rgb}{0.5, 0.09, 0.09}
\usepackage{smartdiagram}
\tikzstyle{decision} = [diamond, draw, 
    text width=1.8cm, text badly centered, node distance=3cm, inner sep=0pt]
\tikzstyle{block} = [rectangle, draw, 
    text width=3.5cm, text centered, rounded corners, minimum height=.5cm]
\tikzstyle{line} = [draw, -latex']
\tikzstyle{cloud} = [draw, ellipse, node distance=2cm,
    minimum height=2em]

\tikzset{
  main/.style={circle, minimum size = 5mm, thick, draw =black!80, node distance = 10mm},
  connect/.style={-latex, thick},
  box/.style={rectangle, draw=black!100}
}

\makeatletter
\DeclareRobustCommand{\rvdots}{%
  \vbox{
    \baselineskip4\p@\lineskiplimit\z@
    \kern-\p@
    \hbox{.}\hbox{.}\hbox{.}
  }}
\makeatother

\newcommand\abs[1]{\left|#1\right|}
\newcommand{\subfigANDtitle}[2][.2\linewidth]{%
  \begin{tabular}{@{}>{\centering\arraybackslash}p{#1}@{}} #2 \end{tabular}}

\begin{document}

\title{A Fully Bayesian Infinite Generative Model for Dynamic Texture Segmentation 
}


\author{Sahar Yousefi         \\
        M. T. Manzuri Shalmani \\
        Antoni B. Chan
}


\institute{Sahar Yousefi  \at
              Sharif University of Technology, Tehran, Iran \\
              \email{syousefi@ce.sharif.edu}           
           \and
            M. T. Manzuri Shalmani \at
              Sharif University of Technology, Tehran, Iran\\
              \email{manzuri@sharif.edu}\\
            \and
            Antoni B. Chan \at
            City University of Hong Kong, Hong Kong \\
            \email{abchan@cityu.edu.hk}
}

\date{Received: date / Accepted: date}

\maketitle

\begin{abstract}
Generative dynamic texture models (GDTMs) are widely used for dynamic texture (DT) segmentation in the video sequences.  GDTMs represent DTs as a set of linear dynamical systems (LDSs). A major limitation of these models concerns the automatic selection of a proper number of DTs. Dirichlet process mixture (DPM) models which have appeared recently as the cornerstone of the non-parametric Bayesian statistics, is an optimistic candidate toward resolving this issue. Under this motivation to resolve the aforementioned drawback, we propose a novel non-parametric fully Bayesian approach for DT segmentation, formulated on the basis of a joint DPM and GDTM construction. This interaction causes the algorithm to overcome the problem of automatic segmentation properly. We derive the Variational Bayesian Expectation-Maximization (VBEM) inference for the proposed model. Moreover, in the E-step of inference, we apply Rauch-Tung-Striebel smoother (RTSS) algorithm on Variational Bayesian LDSs. Ultimately, experiments on different video sequences are performed. Experiment results indicate that the proposed algorithm outperforms the previous methods in efficiency and accuracy noticeably.
\end{abstract}

\section{Introduction}
\label{intro}
Dynamic Texture (DT) segmentation has received considerable attention over the past decade \cite{ghoreyshi2007segmenting}-\cite{zhang2015unsupervised} . DTs are composed of ensembles of particles subject to their stochastic motion. In fact, DTs represent important characteristics of video sequences which have stationary properties in appearance, i.e. spatial domain, and motion, i.e. time domain. 
From the appearance point of view, textures can be divided into: structural and stochastic. The former is mostly artificial and periodic and can be described by Texton, the putative units of pre-attentive human texture perception \cite{julesz1981textons}, such as Tartan, Chessboard, etc. On the contrary, the latter is natural and quasi-periodic such as ocean waves, grass field, etc. In the literature, DTs from the perspective of their particles are divided into three categories: (i) microscopic, such as plumes of smoke, water flow, (ii) macroscopic, such as blowing leaves in wind, (iii) objects such as crowd of people, vehicles in traffic jams \cite{chan2008modeling}. 
Since DT Segmentation relies on both appearance and motion changes, effective texture segmentation in videos is one of the most complicated issues in studies of video processing. There is a great deal of efforts ongoing for this purpose \cite{chen2013automatic}, \cite{badrinarayanan2013semi}-\cite{mumtaz2014joint}. Although \cite{chetverikov2005brief} mentions five categories of approaches for DTs recognition, the DTs segmentation approaches can be grouped into one of three categories: motion-based methods \cite{chen2013automatic}, \cite{vidal2005optical}, \cite{amiaz2007detecting}, spatiotemporal feature based methods \cite{ge2016dynamic}, \cite{chen2008unsupervised}, \cite{gonccalves2013dynamic}, \cite{ojala2002multiresolution} and generative dynamic texture models (GDTMs) \cite{chan2008modeling}, \cite{chan2009layered}, \cite{doretto2003dynamic}. 

The motion-based methods are one of the most commonly used approaches for DT segmentation. In these approaches, DTs are represented by estimating a sequence of motion patterns \cite{gonccalves2013dynamic}, where optical flows are frequently used for this purpose. Amiaz et al. presented an optical flow based method in order to segment videos into static and dynamic texture regions \cite{amiaz2007detecting}. This approach uses a variant of the brightness constancy assumption. After that, they expanded their work by using gradient constancy and color constancy assumptions for the goal \cite{fazekas2009dynamic}. These approaches detect dynamic regions but do not consider any difference between co-occurring DTs. Vidal and Ravichandran proposed a three-steps motion-based method for DT modeling and segmentation \cite{vidal2005optical}. While good results can be achieved in these approaches, accurate motion analysis itself is a challenging task due to the difficulties raised by aperture problem, occlusion, and video noise. Moreover, the optical flow methods are based on the well-known brightness constancy assumption \cite{gennert1987relaxing}. Hence, any variation in the lighting within the scene violates the brightness constancy constraint.

Gon\c{c}alves and Bruno proposed a memory-based approach using partially self-avoiding walks on three orthogonal planes \cite{gonccalves2013dynamic}, in which the segmentation is performed by clustering appearance and motion features. In this approach, the features extraction process is illumination-sensitive, memory-consuming and suffers from heavy computation burden. 

Chen et al. proposed an approach based on local descriptors and optical flows \cite{chen2013automatic},\cite{chen2008unsupervised}. In this method, for considering appearance mode, spatial texture descriptor, i.e. local binary pattern (LBP) and Weber local descriptor (WLD), and for motion mode optical flow and local temporal texture descriptor are used. The LBP \cite{ojala2002multiresolution} is a gray-scale invariant texture descriptor which describes the texture with coding the Textons into binary patterns. Chen et al. proposed volume based LBP, in which for each voxel, a binary code is produced by thresholding its neighborhood with the value of the center voxel. Using the extracted features, the method performs the segmentation by hierarchical splitting, agglomerative merging and pixel-wise classification. Hence, the segmentation results exhibit obvious jaggedness of the boundaries. Furthermore, LBPs are sensitive to noise and illumination variations. Since LBP describes the Textons, it assumes that the textures are structural and cannot achieve a proper result for stochastic textures with the quasi-periodic components, such as ocean waves. Moreover, since different dynamic textures are composed of varying sizes of Textons, finding the best neighbor cube size is a challenging issue. On the other hand, as the neighbor cube gets much larger, more binary patterns are needed. 

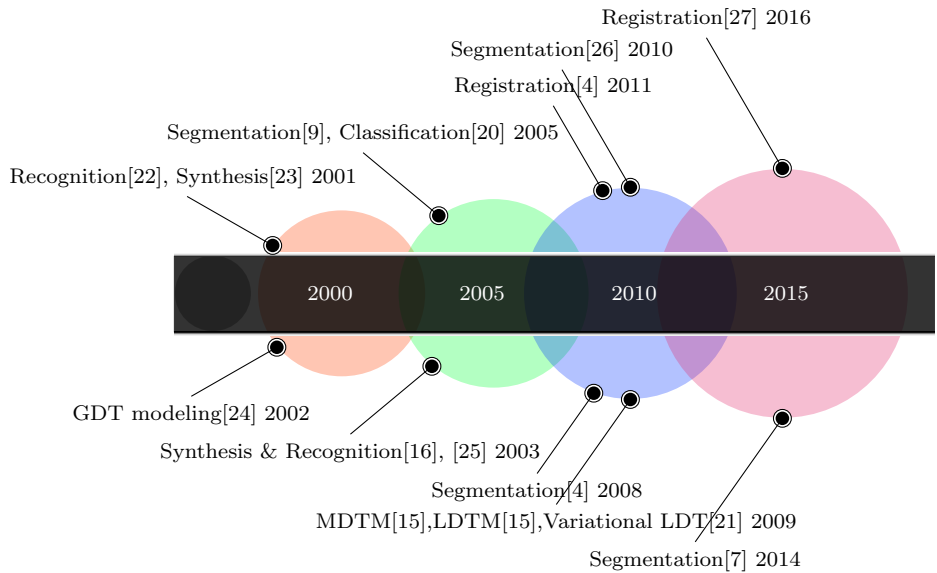
\begin{figure*}
    \begin{tikzpicture}[timespan={}]
    
    \timeline[custom interval=true]{2000,2005,2010,2015}
    
    \begin{phases}
    \initialphase{involvement degree=1.00cm,phase color=black}
    \phase{between week=1 and 2 in 0.1,involvement degree=2.2cm}
    \phase{between week=1 and 2 in 1.1,involvement degree=2.5cm,color=green!80!cyan}
    \phase{between week=2 and 3 in 1.0,involvement degree=2.8cm,phase color=blue!80!cyan}
    \phase{between week=3 and 4 in 1.0,involvement degree=3.3cm,phase color=richmaroon}
    \end{phases}
    
    \addmilestone{at=phase-1.145,direction=150:1.3cm,text={\text{Recognition\cite{saisan2001dynamic}, Synthesis\cite{soatto2001dynamic} 2001 } },text options={above}}
    \addmilestone{at=phase-1.220,direction=210:1.3cm,text={GDT modeling\cite{doretto2002dynamic} 2002 },text options={below}}
    
    \addmilestone{at=phase-2.230,direction=220:1.4cm,text={Synthesis \& Recognition\cite{doretto2003dynamic}, \cite{doretto2003editable} 2003},text options={below}}
    
    \addmilestone{at=phase-2.485,direction=140:1.3cm,text={Segmentation\cite{vidal2005optical}, Classification\cite{chan2005probabilistic} 2005},text options={above}}  
    \addmilestone{at=phase-3.465,direction=120:1.3cm,text={Registration\cite{chan2008modeling} 2011},text options={above}}
    
    \addmilestone{at=phase-3.90,direction=120:1.8cm,text={Segmentation\cite{ravichandran2010unified}} 2010,text options={above}}
    
    \addmilestone{at=phase-3.250,direction=235:1.3cm,text={Segmentation\cite{chan2008modeling} 2008},text options={below}}
    
    \addmilestone{at=phase-3.270,direction=235:1.7cm,text={MDTM\cite{chan2009layered},LDTM\cite{chan2009layered},Variational LDT\cite{chan2009variational} 2009},text options={below}}
    \addmilestone{at=phase-4.90,direction=120:2cm,text={Registration\cite{yang2016dynamic} 2016},text options={above}}
    
    \addmilestone{at=phase-4.270,direction=235:2cm,text={Segmentation\cite{mumtaz2014joint} 2014},text options={below}}

    \end{tikzpicture}
\caption{History for Generative Dynamic Texture Models}
\label{fig:0}       
\end{figure*}

In the literature, GDTMs are introduced as a convenient mean for DT analysis. Figure \ref{fig:0} categorizes some of the recent GDT models for different applications. There are a wide variety of works which consider GDTMs for DT segmentation \cite{chan2009layered}, \cite{mumtaz2014joint}, \cite{doretto2003dynamic}, \cite{rivera2015spatiotemporal}, \cite{chan2005probabilistic}. The GDTM poses DTs as a stochastic visual process over time and space. In these approaches, DTs are modeled by linear dynamical systems (LDSs) \cite{doretto2003dynamic}. Demonstrating a wide variety of complex patterns of motion and spatiotemporal model, GDTMs are able to reach an admissible solution which can model complex senses \cite{chan2008modeling}, \cite{chan2005probabilistic}. The GDTM suffers from the restrictive assumption that the video sequences do not contain co-occurring DTs. For addressing this limitation, various extensions of GDTMs have been proposed \cite{ghoreyshi2007segmenting}, \cite{chan2009layered}. Dynamic texture mixture (DTM) \cite{chan2008modeling} and layered Dynamic Texture (LDT) \cite{chan2009layered} are two of these efforts. The DTM supposes that a collection of spatiotemporal video patches are modeled as samples from a set of underlying dynamic textures. As DTs are globally homogenous and locally inhomogeneous, patch-based approaches lead to poor results. Moreover, the DTM, like all clustering models, is not a global generative model for video of co-occurring textures \cite{chan2009variational}.On the contrary, LDT is a global generative model which supposes videos are modeled as a superposition of DTs. In this method, in order to estimate the parameters of the model, Expectation-Maximization (EM) algorithm for maximum-likelihood is used. Moreover, Gibbs sampler is used for inference. The DTM uses an initial partition and LDT is initialized by the results of the DTM segmentation. \par

To our knowledge, despite the promising results achieved by the recent methods, all of them need an important prerequisite to segment the DTs: determination of the optimal number of DTs in video sequences. As the different frames of video sequences may contain a different number of DTs, using expert knowledge for deriving region segments may be an important restriction for systematic approaches. In this paper, a non-parametric fully Bayesian generative approach based on Dirichlet process (DP) will be introduced for addressing this limitation. DP is a distribution over distributions which commonly used as a prior on the parameters of the mixture model with countably infinite components. In other words, DP is an infinite mixture of distribution with given parametric distribution \cite{ferguson1973bayesian}. Although the model is defined for infinite mixtures, the inference is tractable because parameters of a finite set of mixtures are needed to be determined. For this purpose, Monte Carlo technique \cite{ghahramani2003bayesian} or Variational Bayesian approximation \cite{beal2003variational} can be used, where the latter outperforms the former in speed.\par
The idea of using DPs to define mixture models with an infinite number of components for image segmentation, infinite hidden Markov random field (iHMRF), has been previously explored in \cite{chatzis2010infinite}. Due to the motion mode in DTs, the iHMRF is inadequate for DT segmentation. Beal et al. proposed a model known as the infinite hidden Markov model (iHMM), in which the number of hidden states of the hidden Markov model is allowed to be countably infinite \cite{beal2002infinite}. Due to the continuity of the hidden states in LDSs, iHMM is not proper for DT segmentation. Beal et al. also, proposed Variational Kalman Smoother for one layer LDSs \cite{beal2001variational}.Under this motivation, we introduce a novel non-parametric generative Bayesian model for DT segmentation. Combining GDTM and DP allows us to introduce a novel model that has the ability to set the number of DTs automatically. In our approach, the prior probabilities of the model are jointly affected by DP and GDTM with countably infinite DTs. For inference, we use Variational Bayesian Expectation Maximization (VBEM) approximation which is facilitated by means of mean-field approximation.  In VBE step, 1st-order and 2nd-order expected values of the hidden states are needed. For this purpose, we perform Rauch-Tung-Striebel smoother (RTSS) on Variational Bayesian LDSs.\par
The two main contributions of this paper are: (i) a novel non-parametric formulation of GDTM based on DP for unsupervised co-occurring DT segmentation which resolves the problem of determining the proper number of DTs. (ii) a fully Bayesian generative model which resolves the sensitivity of the GDTM to the initialization of parameters.\par
This paper is organized as follows: in Section \ref{sec:1}, a brief overview of generative dynamic texture model and Dirichlet process is provided. In section \ref{sec:4} the proposed infinite generative dynamic texture model is introduced and an elegant truncated Variational Bayesian inference algorithm for the model is derived. Section \ref{sec:8} describes implementation of the proposed method comprehensively. The evaluation of the efficiency of the proposed model through three types of aforementioned dynamic textures (i.e. microscopic, macroscopic and objects) in comparison with the previous works is presented in section \ref{sec:9}. Finally, the conclusion is obtained in section \ref{sec:10}.

\section{Theoretical Background}
\label{sec:1}
\subsection{Dynamic Texture Model }
\label{sec:2}\par
\begin{figure*}
\centering
\begin{tikzpicture}
\tikzstyle{main}=[circle, minimum size = 5mm, thick, draw =black!80, node distance = 5mm]
\tikzstyle{connect}=[-latex, thick]
\tikzstyle{box}=[rectangle, draw=black!100]
  \node[box,draw=white!100] (Latent) {\textbf{}};
  \node[main,minimum size=1.0cm] (L1) [right=of Latent] {$x_1$};
  \node[main,minimum size=1.0cm] (L2) [right=of L1] {$x_2$};
  \node[main,minimum size=1.0cm] (L3) [right=of L2] {$x_3$};
  \node[main,minimum size=1.0cm] (Lt) [right=of L3] {$x_T$};
  \node[box,draw=white!100] (Observed) [below=of Latent] {\textbf{}};
  \node[main,fill=black,text=white,minimum size=1.0cm] (O1) [right=of Observed,below=of L1] {$y_1$};
  \node[main,fill=black,text=white,minimum size=1.0cm] (O2) [right=of O1,below=of L2] {$y_2$};
  \node[main,fill=black,text=white,minimum size=1.0cm] (O3) [right=of O2,below=of L3] {$y_3$};
  \node[main,fill=black,text=white,minimum size=1.0cm] (Ot) [right=of O3,below=of Lt] {$y_T$};
  \path (L3) -- node[auto=false]{\ldots} (Lt);
  \path (L1) edge [connect] (L2)
        (L2) edge [connect] (L3)
        (L3) -- node[auto=false]{\ldots} (Lt);
        (O3) -- node[auto=false]{\ldots} (Ot);
  \path (L1) edge [connect] (O1);
  \path (L2) edge [connect] (O2);
  \path (L3) edge [connect] (O3);
  \path (Lt) edge [connect] (Ot);
  \draw[dashed]  [below=of L1,above=of O1];
\end{tikzpicture}
\caption{Graphical model of generative dynamic texture model}
\label{fig:1}       
\end{figure*}
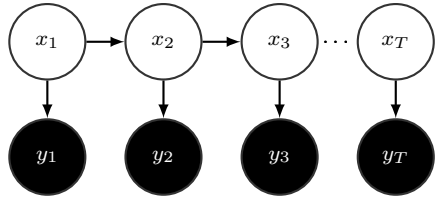

GDTM is a probabilistic approach which models the time-varying texture in the videos using a LDS \cite{soatto2001dynamic}. In this method, a state space model based on auto-regressive moving average (ARMA) models is used. Due to the moving average property, the structure models the non-deterministic data. Moreover, Due to auto-regressive property, the structure models data dependent to the past occurrences. Figure \ref{fig:1} illustrates the graphical model of GDTMs. GDTMs are defined by a set of time-evolving hidden states, $x_t\in\Re^N$ , and a sequence of observations, $y_t\in\Re^M$ , as below

%

\begin{equation}\label{eq:1}
    \left\{
         \begin{array}{ll}
            x_t=Ax_{t-1}+\zeta_t\\
            y_t=C x_t+\xi_t,& t\in[1,T],
         \end{array}
    \right.
\end{equation}
in which $A\in\Re^{N\times N}$ is the transition matrix, $C\in\Re^{M\times N}$ is the observation matrix, $T$ is the temporal length of the video,  $\zeta_t\in\Re^{N}$ and $\xi_t\in\Re^{M}$ are the state noise and the observation noise respectively which are independent and identically distributed (i.i.d.) sequences drawn from a known distribution such as Gaussian. The initial state is distributed from a normal distribution as $x_1\sim N(x_1\mid\delta,S^{-1})$. 
\subsection{Dirichlet Process }
\label{sec:3}
The Dirichlet process prior is a measure on measures \cite{ferguson1973bayesian}. DPs assume that for the random variables $\{\theta_{i}^{*}\}_{i=1}^{L}$, the random measures are drawn from a $DP(G_0,\alpha)$, where $G_0$ is a base distribution and $\alpha$ is a positive scaling parameter i.e. 

\begin{equation}\label{eq:2}
    \begin{array}{ll}
       G\mid\{G_0,\alpha\}\sim DP(G_0,\alpha) \\
       \theta_i^*\mid G\sim G & i=1,...,L.
   \end{array}
\end{equation}

Let $\{\theta_{j}^{*}\}_{j=1}^{K}$  be the set of distinct values taken by variables $\{\theta_{i}^{*}\}_{i=1}^{L-1}$. Denoting as  $f_{j}^{L-1}$ the number of values in $\{\theta_{i}^{*}\}_{i=1}^{L-1}$ that equal to $\theta_j$ , the conditional distribution of  $\theta_L^*$ given $\{\theta_{n}^{*}\}_{n=1}^{\Pi-1}$  has the form 

\begin{equation}\label{eq:3}
    p\left(\theta_L^*\mid \{\theta_{i}^{*}\}_{i=1}^{L-1}, G_0,\alpha\right)=\frac{\alpha}{\alpha+L-1}G_0+\sum_{j=1}^{K}\frac{f_j^{L-1}}{\alpha+L-1}\delta_{\theta_{j}}
\end{equation}
in which $\delta_{\theta_j}$  denotes the distribution concentrated at a single point $\theta_j$ \cite{blackwell1973ferguson}. The parameter $\alpha$  is a tradeoff between sampling a new parameter from the base distribution and sharing a previously sampled parameter. 

Therefore $\alpha$ indicates a key role in determining the number of distinct parameters. As $\alpha$  gets larger and larger $G$  converges to $G_0$. On the contrary, as $\alpha$ gets smaller and smaller, all the values $\{\theta_{i}^{*}\}_{i=1}^{L}$ tend to a single random variable.

Drawing samples from DP can be regarded in terms of stick-breaking construction \cite{sethuraman1994constructive}. A stick breaking prior on the space   has the form 

\begin{equation}\label{eq:4}
    G=\sum_{j=1}^{\infty}\pi_j(\nu)\delta_{\theta_j},
\end{equation}
where 
\begin{equation}\label{eq:5}
   \pi_j(\nu)=\nu_j\prod_{j'=1}^{j-1}\left(1-\nu_{j'}\right) , \pi_j(\nu)\in[0,1],
\end{equation}
and
\begin{equation}\label{eq:6}
   \sum_{j=1}^{\infty}\pi_j(\nu)=1,
\end{equation}
in which $\{\nu_j\}_{j=1}^{\infty}$ and $\{\theta_j\}_{j=1}^{\infty}$  are the random variables drawn from $\nu_j\overset{i.i.d.}{\sim} Beta(1,\alpha)$  and $\theta_j\overset{i.i.d.}{\sim} G_0$ respectively. Suppose $y=\{y_i\}_{i=1}^L$ be the set of observations which is modeled by DP. Each observation $y_i$ is assumed to be drawn from its relative conditional probability density function $p\left(y_i\mid \theta_j^*\right)$ which is parametrized by $\theta_j$. Introducing a discrete random variable $z=\{z_i\}_{i=1}^L$, in which $z_i=j$  denotes that $y_i$ is drawn from $\theta_j^*$, Dirichlet process mixture model with DP can be defined as 

\begin{equation}\label{eq:7}
    \begin{array}{ll}
       y_i\mid z_i=j; \theta_j \sim p(y_i\mid \theta_j),\\
       z_i\mid\pi(\nu)\sim Mult(\pi(\nu)),\\
       \nu_j\mid\alpha\sim Beta(1,\alpha),\\
       \theta_j\mid G_0\sim G_0,
   \end{array}
\end{equation}

in which $\pi(\nu)=\left(\pi_j(\nu)\right)_{j=1}^{\infty}$  is given by (\ref{eq:7}), and  $Mult(.)$ denotes Multinomial distribution. 
\section{Infinite Generative Dynamic Texture Model (IGDTM)}
\label{sec:4}

In this section, we propose a novel probabilistic model named infinite generative dynamic texture model (IGDTM). In this model, it is supposed that the video sequences are composed of infinite number of DTs and each DT can be modeled by a LDS. In other words, the model considers videos as a superposition of the output of countable infinite disjoint GDTMs. In this approach the fully Bayesian reasoning is used for reasoning which endeavors to estimate parameters of an underlying distribution based on the observed distribution. This requires us to specify prior distributions on parameters. For this purpose, the exponential families are used. 

\begin{figure*}
\centering
\begin{tikzpicture}[->,>=stealth',auto,node distance=3cm,
  thick,main node/.style={circle,draw,font=\sffamily\Large\bfseries}]

  \node[box,draw=white!100] (Latent) {\textbf{}};
  \node[main] (X1) [right=.7cm of Latent,minimum size=1.0cm] {$x_1^{(j)}$};
  \node[main] (X2) [right=.7cm of X1,minimum size=1.0cm] {$x_2^{(j)}$};
  \node[main] (X3) [right=.7cm of X2,minimum size=1.0cm] {$x_3^{(j)}$};
  \node[main] (Xt) [right=.7cm of X3,minimum size=1.0cm] {$x_T^{(j)}$};
  \node[main,fill=black,text=white] (Y1) [below=2.5cm of X1,minimum size=1.0cm] {$y_{i1}$};
  \node[main,fill=black,text=white] (Y2) [below=2.5cm of X2,minimum size=1.0cm] {$y_{i2}$};
  \node[main,fill=black,text=white] (Y3) [below=2.5cm of X3,minimum size=1.0cm] {$y_{i3}$};
  \node[main,fill=black,text=white] (Yt) [below=2.5cm of Xt,minimum size=1.0cm] {$y_{iT}$};

  \node[box,draw=white!100,left=of Y1] (Observed) {\textbf{}};
  \path (X3) -- node[auto=false]{\ldots} (Xt);

  \edge {X1} {X2} 
  \edge {X2} {X3}

  \edge {X1} {Y1} 
  \edge {X2} {Y2} 
  \edge {X3} {Y3} 
  \edge {Xt} {Yt} 
   
   \node[main,fill=white] (deltaS) [above=0.5cm of X1,minimum size=1.0cm] {$\delta_j,S_j$};
  \node[main,fill=white] (Q) [above=0.5cm of X2,minimum size=1.0cm] {$Q_j$};
  \node[main,fill=white] (A) [right=0.7cm of Q,minimum size=1.0cm] {$A^{(j)}$};
  \node[main,fill=white] (muR) [left=0.5cm of X1,minimum size=1.0cm] {$\mu_j,r_j$};
  \node[main,fill=white] (Z) [below =of Y2,minimum size=1.0cm] {$Z$};
  \node[main,fill=white] (pi) [below=.3cm of muR,minimum size=1.0cm] {$\pi_j$};
  \node[main,fill=white] (alpha) [left=.5cm of pi,minimum size=1.0cm] {$\alpha$};
  
  \node[main,fill=black,minimum size=.1cm,label={\small $\eta_1,\eta_2$}] (c1) [above=of alpha]{};
  \edge {c1} {alpha}

  
  \node[main,fill=black,minimum size=.1cm,label={\small $w_2,\lambda_2,w_2,\Psi_2$}] (c3) [above=2.3cm of muR]{};
  \edge {c3} {muR}  
  
  \node[main,fill=black,minimum size=1mm,label={\small $m_3,\lambda_3,w_3,\Psi_3$}] (c4) [above=1.1cm of deltaS]{};
  \edge {c4} {deltaS}
  
  \node[main,fill=black,minimum size=.1cm,label={\small $w_1,\Psi_1$}] (c5) [above=1.1cm of Q]{};
  \edge {c5} {Q}
  
  \node[main,fill=black,minimum size=.1cm,label={\small $\sigma_{A}$}] (c6) [above=1.1cm of A]{};
  \edge {c6} {A}
  
  \edge {alpha} {pi} 

   \path[every node/.style={font=\sffamily\small}]
    (pi) edge[bend right] node [left] {} (Z);
    
 \node[main,fill=white,text=black] (C) [below right=.7cm and .5cm of Xt,minimum size=1.0cm] {$C_{i}^{(j)}$};

  \edge {muR} {Y1,Y2,Y3,Yt} 
  \edge {Q} {X2,X3,Xt}
  \edge {A} {X2,X3,Xt}
  \edge {C} {Y1,Y2,Y3,Yt} 
  \edge {deltaS} {X1} 
  
  \node[main,fill=black,minimum size=.1cm,label={\small $\Sigma_C$}] (c7) [right=1.1cm of C]{};
  \edge {c7} {C}
  
  \edge {Z} {Y1,Y2,Y3,Yt}

  \node[main,fill=black,minimum size=.1cm,label={\small $\eta_1,\eta_2$}] (c1) [above=of alpha]{};
  \edge {c1} {alpha}
  
    \plate [inner sep=.3cm,xshift=.03cm,yshift=.3cm] {plate1} { (pi)  (deltaS) (Q) (muR) (X1) (X2) (X3) (Xt) (C) } {$\infty$ }; %
    \plate [inner sep=.1cm,xshift=.001cm,yshift=.01cm] {plate2} { (Y1) (Y2) (Y3) (Yt) (C)} {L }; %
\end{tikzpicture}
\caption{Graphical model of IGDTM}
\label{fig:2}       
\end{figure*}
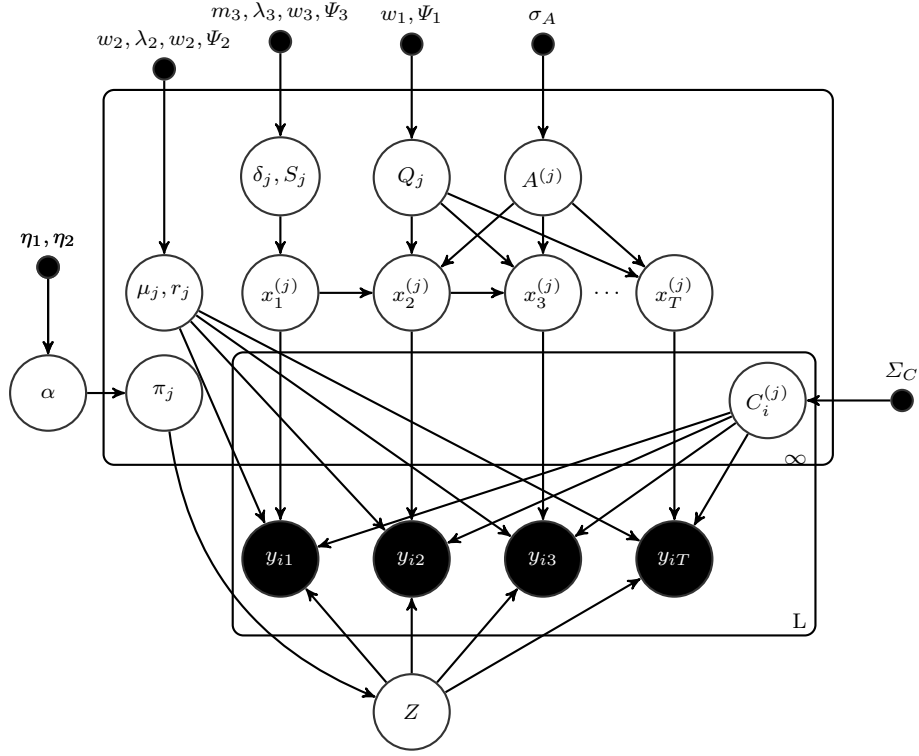

The graphical model of the proposed method illustrated in figure \ref{fig:2}. In this model, the $j^{th}$ DT is modeled by a separate LDS contains a set of hidden states, $x^{(j)}=\{x_t^{(j)}\mid x_t^{(j)}\in \Re^N\}_{t=1}^{T}$ , in which  $T$  is the temporal length of the video. Moreover, The model contains a set of observed variables, $y=\{y_t \mid y_t\in \Re^M , y_t=\{y_{it}\}_{i=1}^L\}_{t=1}^{T}$ (where $y_{it}$ determines the $i^{th}$ pixel, $i\in[1,L]$, on the $t^{th}$ frame, $t\in[1,T]$), and a lattice of sites, $z=\{z_i\}_{i=1}^L$, which represents a Markov random field. The linear dynamical equations of IGDTM are as

\begin{equation}\label{eq:8}
    \left\{
         \begin{array}{ll}
            x_t^{(j)}=A^{(j)}x_{t-1}^{(j)}+\zeta_t, & j\in [1,\infty)\\
            y_{it}=C_{i}^{z_i} x_t^{z_i}+\xi_t^{z_i}, & t\in[1,T],
         \end{array}
    \right.
\end{equation}
where $A^{(j)}$ is the $N\times N$ state transition matrix where
\begin{equation}\label{eq:8_1}
    A^{(j)}=\begin{bmatrix}
        a_{11} & \dots & a_{1N} \\
        \vdots & \ddots & \vdots\\
        a_{N1}&\dots& a_{N N}
    \end{bmatrix},
\end{equation}
in which $a_{n n^{\prime}}\overset{iid}{\sim} N(0,\sigma_A),$ for $n,n^\prime\in[1,N]$, and $C^{(j)}$ is the $L\times N$ observation matrix where
\begin{equation}\label{eq:8_2}
    C^{z_i=j}=\begin{bmatrix}
    c_{11} & \dots & c_{1N}\\
    &\vdots&\\
    c_{L1} & \dots & c_{LN}
    \end{bmatrix},
\end{equation}
where for $i-$th row $c_{ln}\overset{iid}{\sim} N(0,\Sigma_C)$.

The initial state of the $j^{th}$ LDS is as $x_1^{(j)}\sim N(x_1^{(j)}\mid \delta_j ,S_j^{-1})$, in which $\delta_j\in\Re^N$ and $S_j\in\aleph_{+}^N$ are the mean vector and the precision matrix of the $j^{th}$ initial state of the  LDS respectively ( $\aleph_+^N$ denotes all symmetric positive definite $N\times N$  matrices).

Moreover, the state noise process of the $j^{th}$ LDS in the model is given by 

\begin{equation}\label{eq:9}
     \zeta_t\sim N(\zeta_t\mid 0,Q_j^{-1}),
\end{equation}
where $Q_j\in\aleph_+^N$ is the precision matrix of the state noise. Furthermore, the observation noise process of the $j^{th}$ LDS in the model is given by

\begin{equation}\label{eq:10}
     \xi_t^{z_i}\sim N(\xi_t^{z_i}\mid 0,r_j^{-1}),
\end{equation}
where $r_j$ is the precision of the observation noise.  

As IGDTM is the superposition of infinite disjoint DTs, understanding the videos is required to understanding the DTs. The sequence of states in the $j^{th}$ LDS in IGDTM is a Gauss-Markov process which is defined as 

\begin{equation}\label{eq:11}
     p(x^{(j)})= p(x_1^{(j)})\prod_{t=2}^{T} p(x_t^{(j)}\mid x_{t-1}^{(j)}),
\end{equation}
in which the distribution of each state is defined by

\begin{equation}\label{eq:12}
    x_t^{(j)}|x_{t-1}^{(j)},A^{(j)},Q_j\sim N(x_t^{(j)}\mid A^{(j)}x_{t-1}^{(j)},Q_j^{-1}).
\end{equation}

The distribution of the observations is defined as below:

\begin{equation}\label{eq:13}
        y_{it}\mid x_t^{(j)},z_i=j,r_j,\mu_j\sim N(y_{it}\mid C_i^{(j)} x_t^{(j)}+\mu_j,r_j^{-1}),
\end{equation}
where $\mu_j\in\Re$ is the mean of the $j^{th}$ observations of the  LDS. 

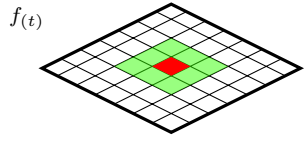
\begin{figure*}
    \centering
    \begin{tikzpicture}[scale=.5,every node/.style={minimum size=1cm},on grid]

    \begin{scope}[
    	yshift=0,every node/.append style={
    	    yslant=0.5,xslant=-1},yslant=0.5,xslant=-1
    	             ]
        \fill[white,fill opacity=.9] (0,0) rectangle (3.4,3.4);
        \draw[black,very thick] (0,0) rectangle (3.4,3.4);
        \draw[step=5mm, black] (0,0) grid (3.4,3.4);
        \fill[green!80!yellow,fill opacity=.5] (1,1) rectangle (2.5,2.5);
        \fill[red] (1.5,1.5) rectangle (2,2);
    \end{scope}
    	             ]
            ]
    \fill[black,font=\footnotesize](-3.8,2) node [above] {$f_{(t)}$};
\end{tikzpicture}
    \caption{Markov neighbourhood system}
    \label{fig:MRF}
\end{figure*}

The distribution of the sites on the label field, $z_i$, are defined by 

\begin{equation}\label{eq:14}
   z_i\mid\hat{z}_{\delta_i},\pi(\nu)\sim p(z_i=j,\mid \tilde{z}_{\delta_i},\beta)p(z_i=j\mid\pi(\nu)),
\end{equation} 
while $p(z_i=j\mid \tilde{z}_{\delta_i},\beta)$ is defined as 
\begin{equation}\label{eq:14_1}
   p(z_i=j\mid\tilde{z}_{\delta_i},\beta)=\frac{-\sum_{i\in c}V_c(\tilde{z}_{ij}\mid\beta)}{\sum_{j^\prime=1}^{K}\exp(-\sum_{i\in c}V_c(\tilde{z}_{ij^\prime}\mid\beta))},
\end{equation} 
which is the approximate point wise probabilities of the $\{z_i\}_{i=1}^{L}$ in order to impose MRF. In equation (\ref{eq:14_1}), $\tilde{z}_{ij}\equiv\left(z_i=j\right)$, $\beta$ is the inverse temperature of the model, $V_c(\tilde{z}_{ij}\mid\beta)$ are the clique potentials, $c$ is the a member of the set of the clique included in the neighbourhood system, $\mathcal{C}$,  $\tilde{z}_{\delta_i}$ is the neighbour set of the cite $z_i$, and $V_c(\tilde{z}_{ij}\mid\beta)$ is comprised of the singleton potential function

\begin{equation}\label{eq:14_2}
   V_{i}(z_i)=
   \left\{
         \begin{array}{ll}
            \varsigma_1,& j=1\\
            \vdots\\
            \varsigma_K,& j=K,
         \end{array}
    \right.
\end{equation} 
and the doubleton potential function
\begin{equation}\label{eq:14_3}
   \forall z_{i^\prime}\in\tilde{z}_{\delta_i}: V_{i,i^\prime}(z_i,z_{i^\prime})=
   \left\{
         \begin{array}{ll}
            \gamma_1,& z_i=z_{i^\prime}, z_i, z_{i^\prime}\in f_{t}\\
            \gamma_2,& z_i\neq z_{i^\prime}, z_i, z_{i^\prime}\in f_{t},
         \end{array}
    \right.
\end{equation} 
where $\varsigma_1,\dots,\varsigma_K,\gamma_1,\gamma_2$ are constants and $f_t$ means the $t^{th}$ frame. Figure \ref{fig:MRF} illustrates neighbourhoods systems. In this figure, the the red sites indicate $z_i$, the green sides indicate $z_{i^\prime}\in f_{t}$.

In this approach, we use the fully Bayesian reasoning in which a prior over all the hidden variables and unknown parameters, $p(\Psi)$,  is introduced and the computation of posterior distribution given all the observations, $y$, and hyper-parameters, $\Xi$, (i.e. $p\left(\Psi\mid\Xi,y\right)$) is interested. For simplicity, we use the conjugate priors. In order to perform Bayesian inference, we apply variational approximation which described in the next section. 
\subsection{Inference for IGDTM  }
\label{sec:5}
As mentioned before, because of better scalability in terms of computational burden, VBEM is used. In this section, we discuss this technique. In order to apply Variational inference, a family of variational distributions must be found that approximates the distribution of the infinite-dimensional random measure $G$ in equation (\ref{eq:4}). It can be done by considering the truncated stick-breaking representation.  For this goal, the mixture proportion,$\pi_j(\nu)$ , is supposed to be zero for $j>K$, where $K$ is a fixed integer value \cite{blei2006variational}.

Bayesian inference technique introduces a set of appropriate prior distribution over the parameters of the model. For simplicity \cite{doretto2003dynamic}, in this paper, the conjugate exponential priors are defined. Hence, we impose Wishart distribution over the covariance matrix of the state noise, $Q_j$, as defined in equations (\ref{eq:16}).
\begin{equation}\label{eq:16}
   Q_j\mid w_1,\Psi_1\sim W(Q_j\mid w_1,\Psi_1),
\end{equation}
in which $w_1$  is the mean vectors and $\Psi_1$ is the covariance matrix of $Q_j$. 

Additionally, a joint one-dimensional Normal-Wishart distribution is imposed as the prior of the mean and covariance matrix of the observations, $\mu_j$, $r_j$ , and the mean and covariance of the initial state, $\sigma_j$, $S_j$ as defined in equations (\ref{eq:17}) and (\ref{eq:18}) respectively.

\begin{equation}\label{eq:17}
   \mu_j,r_j\mid m_2,\lambda_2,w_2,\Psi_2\sim NW_1(\mu_j,r_j\mid m_2,\lambda_2,w_2,\Psi_2),
\end{equation}
\begin{equation}\label{eq:18}
   \delta_j,S_j\mid m_3,\lambda_3,w_3,\Psi_3\sim NW_N(\delta_j,S_j\mid m_3,\lambda_3,w_3,\Psi_3),
\end{equation}
in which $m_1, m_3$ are the mean vector, $\lambda_2, \lambda_3, w_2, w_3$ are four real numbers, and $\Psi_2, \Psi_3$ are the scale matrices.

We use Gamma prior for the Scaling parameter in DP, as 
\begin{equation}\label{eq:19}
   \alpha\mid\eta_1,\eta_2\sim Gam\left(\alpha\mid\eta_1,\eta_2\right),
\end{equation}
where $\eta_1, \eta_2$ are the rate parameters.

Let $\Psi=\{x_{1:T},z_i,\nu_j,\alpha,\delta_j,S_j,Q_j,\mu_j,r_j\}\mid_{i=1,j=1}^{L,K}$ be the set of all hidden variables and unknown parameters and $\Xi=\{w_1,\Psi_1,m_2,\lambda_2,w_2,\Psi_2,m_3,\lambda_3,w_3,\\ \Psi_3,\eta_1,\eta_2,\sigma_A^{(j)},\Sigma_C^{(j)}\}$ be the set of all hyper-parameters of the imposed priors. The joint distribution $p(\Psi,y\mid\Xi)$ is defined as:
\begin{multline}\label{eq:20}
       p\left(\Psi,y\mid\Xi\right)= 
       \prod_{t=2}^{T}\prod_{j=1}^K P\left(x_t^{(j)}\mid A^{(j)}x_{t-1}^{(j)},Q_j\right)
       \prod_{j=1}^K P\left(Q_j\mid w_1,\Psi_1\right)\\
       \times\prod_{i=1}^L\prod_{j=1}^K \prod_{t=1}^T P\left(y_{it}\mid C_i^{(j)}x_t^{(j)}+\mu_j,z_i=j,r_j\right)^{\mathbb{I}(z_i=j)} \prod_{j=1}^K P\left(\mu_j,r_j\mid m_2,\lambda_2,w_2,\Psi_2\right)\\
       \times\prod_{j=1}^K P\left(x_1^{(j)}\mid \delta_j,S_j\right)P\left(\delta_j,S_j\mid m_3,\lambda_3,w_3,\Psi_3\right) \\
       \times\prod_{j=1}^{K} P\left(v_j\mid\alpha\right) P(A^{(j)}\mid 0,\sigma_A)P(C^{(j)}\mid 0,\Sigma_C)P(\alpha\mid\eta_1,\eta_2)P\left(Z\right) ,
\end{multline}
in which
\begin{equation}
    P\left(C^{(j)}\mid 0,\Sigma_C\right)=\prod_{i=1}^L N\left(C_i^{(j)^\intercal}\mid 0,\Sigma_C\right),
\end{equation}
and
\begin{equation}\label{eq:20_1}
    P(Z)=\prod_{i=1}^{L}\prod_{j=1}^{K} p(z_i=j\mid \tilde{z}_{\delta_i},\beta)p(z_i=j\mid v(\pi_j)),
\end{equation}
where $p(z_i=j\mid \tilde{z}_{\delta_i},\beta)$ are the point-wise prior probabilities of the MRF, given by \ref{eq:14_1}, and $p(z_i=j\mid v(\pi_j))$ are the prior probabilities of the model states stemming from the imposed DP, given by \ref{eq:5} and \ref{eq:7}.

Variational Bayesian inference is used for approximating intractable integrals arising in Bayesian inference and machine learning. In this method, the actual posterior over the set of all hidden variables and unknown parameters, i.e. $p\left(\Psi\mid\Xi,y\right)$ , is approximated by a Variational distribution, known as $q_\Psi\left(\Psi\right)$. Mean-field variational Bayesian is the most common type of Variational Bayes, which uses the Kullback$-$Leibler divergence (KL$-$divergence) of $p\left(\Psi\mid\Xi,y\right)$ from $q\left(\Psi\right)$ as the dissimilarity function. Under this assumption, the log marginal likelihood of model yields

\begin{equation}\label{eq:21}
   \log p(y)=\ell(q_\Psi)+\text{KL}(q_\Psi\mid\mid p),
\end{equation}
in which 
\begin{equation}\label{eq:22}
   \ell(q_\Psi)=\int q_\Psi(\Psi)\text{log}\frac{p(\Psi\mid\Xi,y)}{q_\Psi(\Psi)}d\Psi,
\end{equation}
where $\text{KL}(.)$ stands for KL-divergence. Since KL divergence is non-negative, $\ell(q_\Psi)$  forms a strict lower bound of the log marginal likelihood and will be equal when $\text{KL}(q_\Psi\mid\mid p)=0$ or $q_\Psi(\Psi)=p(\Psi\mid\Xi,y)$. As log marginal likelihood is constant by maximizing  $\ell(q_\Psi)$, $\text{KL}(q_\Psi\mid\mid p)$ will be minimized. 

As we defined the conjugate exponential priors, the Variational posterior $q_\Psi(\Psi)$ is expected to be the same distribution of $p(\Psi\mid\Xi,y)$\cite{Bishop2006}, therefore it is expected that $q_\Psi(\Psi)$  factorized as below
\begin{multline}\label{eq:23}
       q(\Psi)=q_x(x_{1:T})q_z(z)\left(\prod_{j=1}^{K}q_\nu(\nu_j)\right)q_\alpha(\alpha)\prod_{j=1}^K q_{\delta,S}\left(\delta_j,S_j\right)\\
       \times \prod_{j=1}^K q_Q(Q_j)\prod_{j=1}^K q_{\mu,R}(\mu_j,r_j)q(A)q(C),
\end{multline}
where
\begin{equation}\label{eq:24}
    q_x(x_{1:T})=\prod_{t=1}^T q(x_t),q_z(z)=\prod_{i=1}^L q_z(z_i),
\end{equation}
and
\begin{equation}
    q(A)=\prod_{j=1}^K\prod_{n=1}^Nq(a_n^{(j)}),q(C)=\prod_{j=1}^K\prod_{i=1}^Nq(C_i^{(j)})
\end{equation}

Making a full mean field assumption in which $q_x(x_{1:T})=\prod_{t=1}^{{T}}q_x(x_t)$ loses crucial information about the hidden state chain needed for accurate inference. Thus, we employ RTSS algorithm which computes the
expected statistics of the hidden states in time $\mathcal{O}(T)$.

By replacing $q(\Psi)$ in (\ref{eq:21}), we will have
\begin{multline}\label{eq:25}
        \ell(q_\Psi)=\int d\Psi
         \begin{bmatrix} q_x(x_{1:T})\prod_{i=1}^L q_z(z_i)\prod_{j=1}^{K}q_\nu(\nu_j)q_\alpha(\alpha)  \\
         \times\prod_{j=1}^K q_{\delta,S}(\delta_j,S_j)\prod_{j=1}^K q_Q(Q_j)\\\times\prod_{j=1}^K q_{\mu,r}(\mu_j,r_j)\\
         \prod_{j=1}^K\prod_{n=1}^Nq(a_n^{(j)})\prod_{j=1}^K\prod_{i=1}^Nq(c_i^{(j)})\\
         \times\begin{pmatrix}\ln P(\Psi,y\mid\Xi)-\ln q_x(x_{1:T})\\-\sum_{i=1}^L\ln q_z(z_i)
         -\sum_{j=1}^{K}\ln q_\nu(\nu_j)-\ln q_\alpha(\alpha)\\-\sum_{j=1}^K\ln q_{\delta,S}(\delta_j,S_j)
         -\sum_{j=1}^K\ln q_Q(Q_j)\\-\sum_{j=1}^K \ln q_{\mu,r}(\mu_j,r_j)\\-\sum_{j=1}^K\sum_{n=1}^Nq(a_n^{(j)})-\sum_{j=1}^K\sum_{i=1}^Nq(C_i^{(j)})
         \end{pmatrix}\end{bmatrix}.
\end{multline}

Variational Bayesian can be seen as an extension of the expectation maximization (EM) algorithm which contains VB-E step and VB-M step and successively converges on optimum parameter values \cite{beal2003variational}. In M-Step, by maximizing of $\ell(q_\Psi)$ over each of the factor $q_\Psi(\Psi)$ in turn, holding the other fixed, the Variational posterior distribution $q_\Psi(\Psi)$ is derived. The following section will describe the computation of the approximated Variational distributions in M-step. 

Table \ref{tab:1} and \ref{tab:2} describe the random variables/parameters and hyperparameters of the proposed method in detail, respectively.  

\begin{table}
\centering
\caption{Random variables and Parameters in IGDTM  }
\label{tab:1}       
\begin{tabular}{llll}
\hline\noalign{\smallskip}
Random variable/&Name & Dimension&Distributed from\\
Parameter& & &\\
\noalign{\smallskip}\hline\noalign{\smallskip}
$x_t^{(j)}$ &States& $x_t^{(j)}\in\Re^N$ & $N\left(x_t^{(j)}\mid A^{(j)}x_{t-1}^{(j)},Q_j^{-1}\right)$ \\
$\mu_j,r_j$ &Covariance matrix of & $r_j\in \Re$ & $NW_1\left(\mu_j,r_j\mid m_2,\lambda_2,w_2,\Psi_2\right)$ \\
&the state noise&&\\
$Q_j$ & Covariance matrix of & $Q_j\in\aleph_+^N$ & $W_N(Q_j\mid w_1,\Psi_1)$ \\
&the observation noise&&\\
&matrix of observations&&\\
$\delta_j, S_j$ &Mean and covariance & $\delta_j\in\Re^N, S_j\in\aleph_+^N$ & $NW_N\left(\delta_j, S_j\mid m_3,\lambda_3,w_3,\Psi_3\right)$ \\
&matrix of the initial state&&\\
$\alpha$ & Scaling parameter in DP & $\alpha\in\Re^K$ & $Gam(\alpha\mid\eta_1,\eta_2)$ \\
$\nu_j$ & Parameter of DP & $\nu_j\in(0,\infty]$ & $Beta(\nu\mid 1,\alpha)$ \\
$\{z_i\}_{i=1}^L$  & Label field & $\{z_i\}\in{1,...,\infty}$ & $Mult(z\mid\pi(\nu))$\\
\noalign{\smallskip}\hline
\end{tabular}
\end{table}

\paragraph{Variational Bayesian M-step (VBM):} 
In this section, the VBM is described, (see Appendix \ref{sec:appA} for derivations). According to the mean field Variational Bayesian, defining $q(z_i=j')=E\left[\mathbb{I}(z_i=j')\right]; j'\in[1,K]$ 



The distribution of covariance matrix of the observation noise is approximated by $q_Q(Q_{j})=W(Q_{j}|\hat{w}_1,\hat{\Psi}_1)$, in which
\begin{equation}\label{eq:29}
    \hat{w}_1=w_1+(T-1)
\end{equation}
\begin{equation}\label{eq:30}
    \hat{\Psi_1}^{-1}=\Psi_1^{-1}+\sum_{t=2}^T E_{\backslash Q_j}\left[\left(x_t^{(j)}-A^{(j)}x_{t-1}^{(j)}\right)\left(x_t^{(j)}-A^{(j)}x_{t-1}^{(j)}\right)^\intercal\right]
\end{equation}

The distribution of Mean and covariance matrix of observations is approximated by
\begin{equation}\label{eq:31}
    q_{\mu,r}\left(\mu_{j},r_{j}\right)=NW\left(\hat{m}_2,\hat{\lambda}_2,\hat{w}_2,\hat{\Psi}_2\right)
\end{equation}
in which, according to the below notations
\begin{equation}\label{eq:32}
    N_{j}^{(2)}=\sum_{i=1}^L q(z_i=j),
\end{equation}
\begin{equation}\label{eq:33}
    \bar{y}_{tj}=\frac{1}{N_{j}^{(2)}}\sum_{i=1}^L q(z_i=j)y_{it},
\end{equation}
\begin{table}
\centering
\caption{Hyper Parameters in IGDTM}
\label{tab:2}       
\begin{tabular}{llll}
\hline\noalign{\smallskip}
Hyper Parameter & Name & Random Variable/ & Properties/ \\
& & Parameter & Dimension\\
\noalign{\smallskip}\hline\noalign{\smallskip}
$w_1$ & Degree of freedom  & $Q_j$ & $w_1 > N-1$\\
$\Psi_1$ & Scale matrix& $Q_j$ & $\Psi_1 \in \aleph_+^N$ \\
\\
$m_2$ & Mean & $\mu_j$ & $m_2\in\Re$\\
$\lambda_2$& A real number & $\mu_j$ & $\lambda_2>0$\\
$w_2$ & A real number & $r_j$ & $w_2>0$\\
$\Psi_21$& Scale matrix & $r_j$ & $\Psi_2\in\Re$\\
\\

$m_3$ & Mean & $\delta_j$ & $m_3\in\Re^N$\\
$\lambda_3$& A real number & $\delta_j$ &$\lambda_3>0$ \\
$w_3$& A real number & $S_j$ &$w_3>N-1$ \\
$\Psi_3$& Scale matrix& $S_j$ &$\Psi_3\in\aleph_+^N$ \\
\\
$\eta_1$ & Shape parameter & $\alpha$ & $\eta_1>0$ \\
$\eta_2$ & Rate parameter & $\alpha$ & $\eta_2>0$\\
\\
$\sigma_A^{(j)}$ & variance & $A$ & $\sigma_A^{(j)}\in\Re$\\
$\Sigma_C^{(j)}$ & covariance matrix & $C_i^{(j)^\intercal};i\in[1,L]$ & $\Sigma_C^{(j)}\in\aleph_+^N$\\
\noalign{\smallskip}\hline
\end{tabular}
\end{table}
we will have
\begin{equation}\label{eq:35}
    \begin{array}{cc}
        \hat{w}_2=w_2+TN_{j}^{(2)},
    \end{array}
\end{equation}
\begin{equation}\label{eq:36}
    \begin{array}{cc}
        \hat{\lambda}_2=\lambda_2+TN_{j}^{(2)},
    \end{array}
\end{equation}
\begin{equation}\label{eq:37}
    \begin{array}{cc}
        \hat{m}_2=\frac{\lambda_2 m_2+N_{j}^{(2)}\sum_{t=1}^{T}\bar{y}_{tj}}{\hat{\lambda}_2},
    \end{array}
\end{equation}
\begin{equation}\label{eq:38}
    \begin{array}{cc}
        \hat{\Psi}_2^{-1}=\Psi_2^{-1}-\\\frac{1}{\hat{\lambda}_2}\left(\lambda_2^2m_2 m_2^\intercal+\sum_{t=1}^T\bar{y}_{tj}N_{j}^{(2)}m_2^\intercal\lambda_2+      m_2\lambda_2N_j^{(2)^\intercal}\sum_{t=1}^T\bar{y}_{tj}^\intercal+\sum_{t=1}^T\bar{x}_{tj}N_j^{(2)}N_j^{(2)^\intercal}\sum_{t=1}^T\bar{y}_{tj}^\intercal \right)+ \\\lambda_2 m_2m_2^\intercal+
        \sum_{t=1}^T\sum_{i=1}^Lq(z_i=j)y_{it}y_{it}^\intercal-\sum_{t=1}^T\sum_{i=1}^Lq(z_i=j)\hat{\mu}_{C_i}^{(j)}E[x_t^{(j)}]y_{it}^\intercal-
        \sum_{t=1}^T\sum_{i=1}^Lq(z_i=j)y_{it}E\left[x_t^{(j)^\intercal}\right]\hat{\mu}_{C_i}^{(j)^\intercal} +\\\sum_{t=1}^T\sum_{i=1}^L q(z_i=j)\sum_{n=1}^{N}\sum_{{n}^\prime}^{N}       \left({\hat{\Sigma}_{C_{i{n{n}^\prime}}}^{(j)}}
        +\mu_{{C}_{in}}^{(j)}\mu_{C_{i{n^\prime=1}}}^{(j)}
        \right)E\left[x_{t{n^\prime}}^{(j)}x_{t{n}}^{(j)}\right]
    \end{array}
\end{equation}
The distribution of the mean and the covariance matrix of initial state is approximated by

\begin{equation}\label{eq:39}
    q_{\delta,S}(\delta_{j},S_{j})=NW(\hat{m}_3,\hat{\lambda}_3,\hat{w}_3,\hat{\Psi}_3)
\end{equation}
in which, according to the below notations

\begin{equation}\label{eq:40}
    N_{j}^{(3)}=\sum_{i=1}^{L}q(z_i=j),     
\end{equation}
\begin{equation}\label{eq:41}
    \bar{x}_{j}^{(3)}=\frac{1}{N_{j}^{(3)}}\sum_{i=1}^L q(z_i=j)E\left[x_1^{(j)}\right],
\end{equation}
\begin{equation}\label{eq:42}
     \Delta_{j}^{(3)}=\sum_{i=1}^L q(z_i=j)E\left[\left(x_1^{(j)}-\bar{x}_{j}^{(3)}\right)\left(x_1^{(j)}-\bar{x}_{j}^{(3)}\right)^\intercal\right]
\end{equation}
we will have
\begin{equation}\label{eq:43}
        \hat{w}_3=w_3+N_{j}^{(3)}, 
\end{equation}
\begin{equation}\label{eq:44}
        \hat{\lambda}_3=\lambda_3+N_{j}^{(3)},
\end{equation}
\begin{equation}\label{eq:45}
        \hat{m}_3=\frac{\lambda_3 m_3+N_{j}^{(3)}\bar{x}_j^{(3)}}{\hat{\lambda}_3},
\end{equation}
\begin{equation}\label{eq:46}
        \hat{\Psi}_3^{-1}=\Psi_3^{-1}+\Delta_{j}^{(3)}+\frac{\lambda_3 N_{j}^{(3)}}{\hat{\lambda}_3}\left(m_3-\bar{x}_{j}\right)\left(m_3-\bar{x}_{j}\right)^\intercal.
\end{equation}

The distribution of Dirichlet process scaling parameter is approximated by
\begin{equation}\label{eq:47}
    q_\alpha(\alpha)=Gam(\alpha\mid\hat{\eta}_1,\hat{\eta}_2).
\end{equation}
where
\begin{equation}\label{eq:48}
        \hat{\eta}_1=\eta_1,
\end{equation}
\begin{equation}\label{eq:49}
        \hat{\eta}_2=\eta_2-\sum_{j=1}^{K-1}E_{\backslash\alpha}\left[\text{ln}\left(1-\pi_j\right)\right].
\end{equation}

The distribution of Dirichlet process parameter is described by

\begin{equation}\label{eq:50}
    q_\nu\left(\nu_{j}\right)=Beta\left(\nu_{j}|\hat{\beta}_{j,1},\hat{\beta}_{j,2} \right)
\end{equation}
where
\begin{equation}\label{eq:51}
    \hat{\beta}_{j,1}=\sum_{i=1}^{L}q\left(z_i>j^\prime\right)+1,
\end{equation}
\begin{equation}\label{eq:52}
    \hat{\beta}_{j,2}=E\left[\alpha\right]+\sum_{i=1}^L q\left(z_i=j\right).
\end{equation}

The distribution of $q(a^{(j)})$ is described by

\begin{equation}\label{eq:51_1}
    q(A^{(j)})=\prod_{n=1}^N N\left(a^{(j)}_n\mid\hat{\mu}_{A_n}^{(j)},\hat{\sigma}_{A_n}^{(j)}\right)
\end{equation}{}
where
\begin{equation}\label{eq:51_2}
    \hat{\sigma}_{A_n}^{(j)}=\left( \hat{w}_1 \hat{\Psi}_{1_{nn^\prime}}\sum_{t=2}^T E\left[x_{{t-1}_ n}^{(j)^\intercal}x_{{t-1}_ n}^{(j)}\right]+\sigma_A^{(j)}q(n,n) \right)
\end{equation}{}
and
\begin{multline}\label{eq:51_3}
    \hat{\mu}_{A_n}^{(j)}=\hat{\sigma}_A^{(j)^{-1}}\left( \sum_{n^\prime=1}^N \hat{w}_1 \hat{\Psi}_{1_{nn^\prime}} \sum_{t=2}^T E\left[x_{{t-1}_n}^{(j)^\intercal} x_{t_{n^{\prime}}}^{(j)} \right] \right.\\ \left. +\sum_{n^\prime=1}^N \hat{w}_1 \hat{\Psi}_{1_{nn^\prime}} \sum_{t=2}^T E\left[x_{{t-1}_n}^{(j)^\intercal} x_{{t-1}_{n^{\prime}}}^{(j)}\right]a_{n^\prime}^{(j)} +\sum_{n^\prime=1}^N \sigma_A^{(j)}q(n,n^\prime) a_{n^\prime}^{(j)}
    \right)
\end{multline}{}

For $q(C_i^{(j)^\intercal})$

\begin{equation}\label{eq:51_4}
    q\left(C_i^{(j)^\intercal}\right)=N\left(C_i^{(j)^\intercal}\mid\hat{\mu}_{C_i}^{(j)},\hat{\Sigma}_{C_{i}}^{(j)}\right)
\end{equation}{}
where
\begin{equation}\label{eq:51_5}
    \hat{\Sigma}_{C_{i}}^{(j)}=E\left[x_t^{(j)}\right]\left(E\left[r_j\right]y_{it}-E\left[r_j\mu_j\right]\right)
\end{equation}{}
and
\begin{equation}\label{eq:51_6}
    \hat{\mu}_{C_i}^{(j)}= \hat{\Sigma}_{C_{i}}^{(j)^{-1}}\left( \sum_{t=1}^T q(z_i=j) E\left[x_t^{(j)}x_t^{(j)^\intercal}\right] E\left[r_j\right]+\Sigma_{C_i}^{(j)}\right)
\end{equation}{}

Finally, the distribution of label field is approximated by
\begin{multline}\label{eq:53}
    ln q_z\left(z_{i^\prime}=j\right)=\ln p(z_{i^\prime}=j\mid\tilde{z}_{\delta_i},\beta)+\sum_{j=1}^K q\left(z_{i^\prime}=j\right)\\
    \times\begin{pmatrix}E\begin{pmatrix}
    x_1^{(j)^\intercal}S_jx_1^{(j)} -\delta_j^{\intercal}S_jx_1^{(j)}-x_1^{(j)^\intercal}S_j\delta_j+\delta_j^{\intercal}S_j\delta_j
    \end{pmatrix} \\
     +\sum_{t=2}^TE\begin{pmatrix}x_t^{(j)^\intercal}Q_jx_t^{(j)}-x_{t-1}^{(j)^\intercal}A^{(j)^\intercal}Q_jx_t^{(j)}\\
    -x_t^{(j)^\intercal}Q_jA^{(j)}x_{t-1}^{(j)}+x_{t-1}^{(j)^\intercal}A^{(j)^\intercal}Q_jA^{(j)}x_{t-1}^{(j)}\end{pmatrix} \\
    +\sum_{t=1}^TE\begin{pmatrix}y_{i^\prime t}^\intercal R_j y_{it} - x_{t}^{(j)^\intercal}C^{(j)^\intercal}R_j y_{it}\\
    - y_{i^\prime t}R_j C^{(j)}x_t^{(j)}+x_t^{(j)^\intercal}C^{(j)^\intercal}R_jCx_t^{(j)}\end{pmatrix} \\
     +\sum_{t=1}^T E\begin{pmatrix}y_{i^\prime t}^\intercal\Sigma y_{i^\prime t}-\mu_j^\intercal\Sigma_j y_{it}-y_{i^\prime t}^\intercal\Sigma_j\mu_j+\mu_j^\intercal\Sigma_j\mu_j\end{pmatrix} \\
    -E\begin{pmatrix}\ln\abs{S_j}\end{pmatrix} -(T-1)E\begin{pmatrix}\ln\abs{Q_j}\end{pmatrix} \\
    +T E\begin{pmatrix}\ln{\nu_j}\end{pmatrix}-TE\begin{pmatrix}\ln{\abs{R_j}}\end{pmatrix}-TE\begin{pmatrix}\ln{\Sigma_j}\end{pmatrix} \end{pmatrix}\\
    +\sum_{t=1}^T\sum_{j=1}^K q\left(z_{i^\prime}>j\right)E\begin{pmatrix}\ln(1-\nu_j)\end{pmatrix}.
\end{multline}

In each iteration of Variational Bayesian inference, after updating equations (\ref{eq:26})-(\ref{eq:53}), the estimation of $\hat{z}_i$ must be updated as the last step. For this purpose we update $\hat{z}_i$ by maximization of $q_z(z_i=j)$ over $j$ as
\begin{equation}\label{eq:54}
    \hat{z}_i=\arg\max_{j=1:K}q_z(z_i=j).
\end{equation}

In Variational Bayesian E-step, the expected values of the parameters are computed. The equations  of this step will be explained in the next section.

\paragraph{Variational Bayesian E-step (VBE):} 
In this step, the below sufficient statistics are required 

\begin{equation}\label{eq:55}
    \begin{array}{cc}
        E_{q_{Q}}[Q_j]=\hat{w}_1\hat{\Psi}_1, 
    \end{array}
\end{equation}
\begin{equation}\label{eq:56}
    \begin{array}{cc}
        E_{q_{\mu,r}}[r_j]=\hat{w}_2\hat{\Psi}_2,
    \end{array}
\end{equation}
\begin{equation}\label{eq:58}
         E_{q_{\delta,S}}[S_j]=\hat{w}_3\hat{\Psi}_3,
\end{equation}
\begin{equation}\label{eq:59}
        E_{q_{\mu,r}}[r_j\mu_j]=\hat{w}_2\hat{\Psi}_2\hat{m}_2,
\end{equation}
\begin{equation}\label{eq:60}
        E_{q_{\delta,S}}[S_j\delta_j]=\hat{w}_3\hat{\Psi}_3\hat{m}_3,
\end{equation}
\begin{equation}\label{eq:61}
        E_{q_{\alpha}}[\ln{(\alpha)}]=\psi\left(\hat{\eta}_1\right)-\psi\left(\hat{\eta}_2\right),
\end{equation}
\begin{equation}\label{eq:62}
        E_{q_{\pi}}[\ln(1-\pi_j)]=\psi\left(\hat{\eta}_1\right)-\psi\left(\hat{\eta}_1+\hat{\eta}_2\right),
\end{equation}
\begin{equation}\label{eq:63}
        E_{q_{Q}}\left[\ln{\abs{Q_j}}\right]=N\ln{2}+\ln{\abs{\hat{\Psi}_1}}+\sum_{n=1}^N\psi\left(\frac{\hat{w}_1-n+1}{2}\right),
\end{equation}
\begin{equation}\label{eq:64}
        E_{q_{\mu,r}}\left[\ln{\abs{r_j}}\right]=\ln{2}+\ln{\abs{\hat{\Psi}_2}}+\psi\left(\frac{\hat{w}_2}{2}\right), 
\end{equation}

\begin{equation}\label{eq:66}
        E_{q_{\delta,S}}\left[\ln{\abs{S_j}}\right]=N\ln{2}+\ln{\abs{\hat{\Psi}_3}}+\sum_{n=1}^N\psi\left(\frac{\hat{w}_3-n+1}{2}\right),
\end{equation}

\begin{equation}\label{eq:67}
     E_{q_{\nu}}\left[\ln{\nu_j}\right]=\psi\left(\hat{\beta}_{j,1}\right)-\psi\left(\hat{\beta}_{j,1}+\hat{\beta}_{j,2}\right),
\end{equation}
\begin{equation}\label{eq:68}
     E_{q_{\nu}}\left[\ln{(1-\nu_j)}\right]=\psi\left(\hat{\beta}_{j,2}\right)-\psi\left(\hat{\beta}_{j,1}+\hat{\beta}_{j,2}\right),
\end{equation}
where $\psi(.)$ denotes the digamma function. Moreover, we need the expected value of
\begin{equation}\label{eq:69}
    E\left[x_t^{(j)}\right], 
\end{equation}
\begin{equation}\label{eq:70}
    E\left[x_t^{(j)}x_t^{(j)^\intercal}\right], 
\end{equation}
\begin{equation}\label{eq:71}
    E\left[x_t^{(j)}x_{t-1}^{(j)^\intercal}\right].
\end{equation}

For this purpose, we use RTSS algorithm on Variational Bayesian Linear Dynamical system (VBLDS) model, which will be described in the next section. 

\subsection{Rauch-Tung-Striebel smoother (RTSS) algorithm}
\label{sec:6}

In this section, RTSS algorithm for IGDTM is described. RTSS method contains two steps: Forward recursion and backward recursion. In forward recursion we define $\alpha_t(x_t^{(j)})$ to be as the posterior over the hidden states at time $t$ given observed data up to and including time $t$ , i.e.:

\begin{equation}\label{eq:72}
    \alpha_t(x_t^{(j)})\equiv p(x_t^{(j)}\mid\bar{y}_{1:t}^{(j)}),
\end{equation}
in which,
\begin{equation}\label{eq:73}
    \bar{y}_{t^\prime}^{(j)}=\frac{\sum_{i=1}^L q(z_i=j)y_{it^\prime}}{\widetilde{N}_{ij}};t^\prime\in[1,T],j\in[1,K],
\end{equation}
and
\begin{equation}\label{eq:73}
    \widetilde{N}_{ij}={\sum_{i=1}^Lq(z_i=j)}.
\end{equation}
It can be shown that $\alpha_t(x_t^{(j)})=N(x_t^{(j)}\mid\dot{\mu}_t,\dot{\Sigma}_t^{-1})$, in which $\dot{\mu}_t$ and $\dot{\Sigma}_t$ are the mean and covariance matrix respectively and are defined as

\begin{equation}\label{eq:74}
    \dot{\mu}_t=\dot{\Sigma}_t^{-1}\left(\frac{1}{\widetilde{N}_{ij}}\sum_{i=1}^j q(z_i=j) C_i^{(j)^\intercal}r_j(\bar{y}_t^{(j)}-\mu_j)+Q_j^\intercal A^{(j)}\ddot{\Sigma}_{t-1}^{-\intercal}\dot{\Sigma}_{t-1}\dot{\mu}_{t-1}\right),
\end{equation}
\begin{multline}\label{eq:75}
    \dot{\Sigma}_t=\left(Q_j+\frac{1}{\widetilde{N}_{ij}}\sum_{i=1}^j q(z_i=j) C^{(j)^\intercal}R_j \frac{1}{\widetilde{N}_{ij}}\sum_{i=1}^j q(z_i=j) C^{(j)}-\right.\\\left.Q_j^{\intercal}A^{(j)}\ddot{\Sigma}_{t-1}^{-\intercal}A^{(j)^\intercal}Q_j\right),
\end{multline}
in which $^{-\intercal}$  denotes the inverse transpose of matrices, and $\ddot{\Sigma}_t$ is defiend as
\begin{equation}\label{eq:76}
    \ddot{\Sigma}_{t-1}=\left(\dot{\Sigma}_{t-1}+A^{(j)^\intercal}Q_jA^{(j)}\right) 
\end{equation}

The derivations can be found in the Appendix \ref{sec:appB}. In the backward recursion we define $\beta_{t-1}(x_{t-1}^{(j)})$  to be 
\begin{equation}\label{eq:77}
    \beta_{t-1}(x_{t-1}^{(j)})=p(\bar{y}_{t:T}^{(j)}\mid x_{t-1}^{(j)}).
\end{equation}

In the Appendix \ref{sec:appB}, It is shown that $\beta_{t-1}(x_{t-1}^{(j)})=N(x_{t-1}^{(j)}\mid\eta_{t-1},\psi_{t-1}^{-1})$, in which

\begin{equation}\label{eq:78}
    \psi_{t-1}=\left(A^{(j)^\intercal}Q_j A^{(j)}-A^{(j)^\intercal}Q_j^\intercal\psi_t^{\prime^{-\intercal}}Q_j A^{(j)}\right),
\end{equation}
\begin{equation}\label{eq:79}
    \eta_{t-1}=\psi_{t-1}^{-1}A^{(j)^\intercal}Q_j^{\intercal}\psi_t^{\prime^{-\intercal}}\left(\psi_t\eta_t+\frac{1}{\widetilde{N}_{ij}}\sum_{i=1}^L q(z_i=j)C^{(j)^\intercal}_i r_j(\bar{y}_t^{(j)}-\mu_j)\right),
\end{equation}
where $\psi^\prime_t$ is defined as:
\begin{equation}\label{eq:80}
    \psi_t^\prime=\left(Q_j+\psi_t+\frac{1}{\widetilde{N}_{ij}}\sum_{i=1}^L q(z_i=j)C^{(j)^\intercal}_ir_j \frac{1}{\widetilde{N}_{ij}}\sum_{i=1}^L q(z_i=j)C^{(j)}_i\right).
\end{equation}
According to the below notations

It is easy to show that
\begin{equation}\label{eq:81}
    E\left[x_{t}^{(j)}\right]=\omega_t,
\end{equation}
\begin{equation}\label{eq:82}
    E\left[x_{t}^{(j)}x_{t}^{(j)^\intercal}\right]=\Gamma_{t,t}^{-1}+\omega_t\omega_t^\intercal,
\end{equation}
\begin{equation}\label{eq:83}
    E\left[x_{t}^{(j)}x_{t-1}^{(j)^\intercal}\right]=A^{(j)}(\Gamma_{t-1,t-1}^{-1}+\omega_{t-1}\omega_{t-1}^\intercal),
\end{equation}
in which
\begin{equation}\label{eq:84}
    \Gamma_{t,t}=\left(\dot{\Sigma}_t+A^{(j)^\intercal}Q_jA^{(j)}+\psi_t\right),
\end{equation}
\begin{equation}\label{eq:85}
   \Gamma_{t+1,t+1}=\left(Q_j+\frac{1}{\widetilde{N}_{ij}}\sum_{i=1}^L q(z_i=j) C^{(j)^\intercal}_i R_j \frac{1}{\widetilde{N}_{ij}}\sum_{i=1}^L q(z_i=j) C^{(j)}_i\right),
\end{equation}
\begin{equation}\label{eq:86}
    \Gamma_{t+1,t}=\left(Q_jA^{(j)}\right),
\end{equation}
\begin{equation}\label{eq:87}
    \Gamma_{t,t+1}=\left(A^{(j)^\intercal}Q_j\right),
\end{equation}
\begin{equation}\label{eq:88}
    X=\left(\Gamma_{t,t+1}^{-1}\Gamma_{t,t}-\Gamma_{t+1,t+1}^{-1}\Gamma_{t+1,t}\right),
\end{equation}

\begin{multline}\label{eq:89}
    \omega_t=X^{-1}\left(\Gamma_{t,t+1}^{-1}\left(\dot{\Sigma}_t\dot{\mu}_t+\psi_t\eta_t\right)-\right.\\\left.\Gamma_{t+1,t+1}^{-1}\left(\frac{1}{\widetilde{N}_{ij}}\sum_{i=1}^L q(z_i=j)C^{(j)^\intercal}_i r_j\bar{y}_{t+1}^{(j)}\right)\right),
\end{multline}

\begin{multline}\label{eq:90}
    \omega_{t+1}=X^{-1}\left(\Gamma_{t,t}^{-1}\left(\dot{\Sigma}_t\dot{\mu}_t+\psi_t\eta_t\right)-\right.\\\left.\Gamma_{t+1,t}^{-1}\left(\frac{1}{\widetilde{N}_{ij}}\sum_{i=1}^L q(z_i=j) C^{(j)^\intercal}_i r_j\bar{y}_{t+1}^{(j)}\right)\right).
\end{multline}
The derivations can be found in Appendix \ref{sec:appB}.

\section{Implementation of the model}
\label{sec:8}
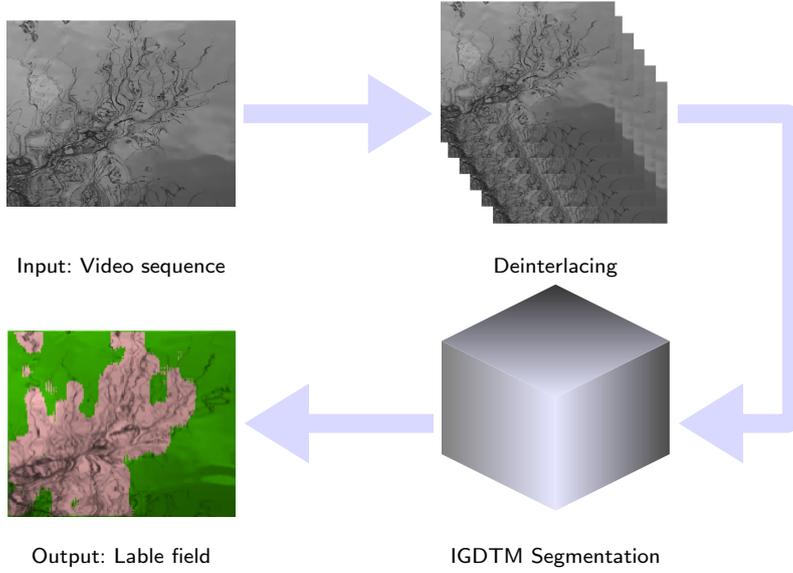
\begin{figure*}
\centering
 \begin{tikzpicture}[
      my text/.style={rounded corners=2pt, text width=50mm, font=\sffamily, line width=.5pt, align=left},
      my arrow/.style={rounded corners=2pt, draw=blue!15, line width=2.5mm, -{Triangle[]}},
      complexnode/.pic={
      \shade[yslant=-0.5,right color=blue!10, left color=black!50]
        (-1.5,-1) rectangle +(1.5,1.5);
      \draw[yslant=-0.5] (-1,-1) ;
      \shade[yslant=0.5,right color=black!70,left color=blue!10]
        (0,-1.0) rectangle +(1.5,1.5);
      \draw[yslant=0.5] (0.5,-1.5) ;
      \shade[yslant=0.5,xslant=-1,bottom color=blue!10,
        top color=black!80] (2,2) rectangle +(-1.5,-1.5);
      \draw[yslant=0.5,xslant=-1] (0.5,0);
      }
    ]
    \node (b1) [my text,align=center] { Input: Video sequence};
    \node (b2) [right=5mm of b1, my text,align=center] {Deinterlacing};
    \node (a2) [above=2mm of b2] {\includegraphics[width=3cm]{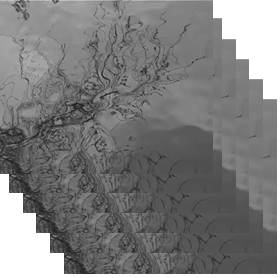}};
    \node (a1) [left=25mm of a2] {\includegraphics[width=3cm]{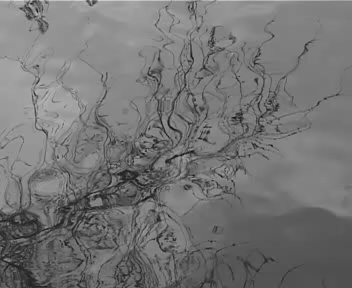}};

    \node (c1) [below=5mm of b2] {\includegraphics[width=30mm]{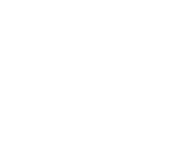}};
    \node (c2) [left=25mm of c1]  {\includegraphics[width=30mm]{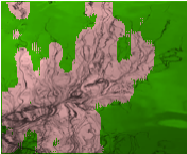}};
    
    \draw  pic [below=2cm of b2] {complexnode} ;
   
    \node (d1) [below=2mm of c1, my text,align=center] {IGDTM Segmentation};
    \node (d2) [below=2mm of c2, my text,align=center] {Output: Lable field};

    \foreach \i/\j in {a1/a2,c1/c2}
    \path [my arrow] (\i) -- (\j);

    \path [my arrow] (a2) -| ($(b2.south east) + (5mm,0)$) |-  (c1);
\end{tikzpicture}
\caption{Process diagram of the IGDTM}
\label{fig:4}       
\end{figure*}

We explained the infinite generative dynamic texture model (IGDTM) in the previous sections. In this section, considering the tables \ref{tab:1} and \ref{tab:2} and computed equations were described previously, we explain the steps of the proposed method. The process diagram of our method is represented in figure \ref{fig:4}. First of all, the input video sequences are de-interlaced using an algorithm such as the spatiotemporal median filter. Then, the IGDTM is applied on the features and the label filed is resulted as the output. 

\begin{figure*}
\centering
\begin{tikzpicture}[node distance = 1.5cm, auto]
    \node [cloud,minimum width=2cm] (start) {Start};
    
    \node [block, below of=start,node distance=1.5cm] (readimg) {Read images and extract observations};
    
    \node [block, below of=readimg] (initialize) {Initialize hyper parameters randomly};
    
    \node [decision,  below of=initialize,node distance=2cm] (converged) {Converged?};

    \node [rectangle, draw, text width=1cm, text centered, rounded corners, minimum height=1cm,  right of= converged,node distance=3cm] (initial) {$j=1$};

    \node [block,  left of= converged,node distance=4cm] (compute) {Compute $\{q_z(z_i=j)\}_{i=1,j=1}^{L,K}$};
    
    \node [rectangle, draw,  text width=4.5cm, text centered, rounded corners, minimum height=1cm,below of= compute,node distance=2cm] (argmax) { $\hat{z}_i=\arg\max_{j=1:K}q_z(z_i=j)$};
    
    \node [decision,below of= initial, minimum size=2cm] (jk) {$j<K?$};
    
    \node [block, below left of=jk,node distance=4.5cm] (VBE) {Variational Baysian Expectation for $j^{th}$ LDS; eq. (\ref{eq:55})-(\ref{eq:68})};
    
    \node [rectangle, draw,     text width=10cm, text centered, rounded corners, minimum height=2cm,below of= VBE,node distance=2cm]  (forwardbackward) {Forward-Backward algorithm for $j^{th}$ LDS\\\begin{enumerate}      \item Compute forward messages; eq. (\ref{eq:72})-(\ref{eq:76})      \item Compute backward messages; eq. (\ref{eq:77})-(\ref{eq:80})       \item Compute $1^{th}$-order and $2^{nd}$-order expectation of states; eq. (\ref{eq:81})-(\ref{eq:83})     \end{enumerate}};
    
    \node [rectangle, draw,  text width=3.5cm, text centered, rounded corners, minimum height=1cm,, below of=forwardbackward,node distance=2cm] (VBM) {Variational Baysian Maximization for $j^{th}$ LDS; eq. (\ref{eq:26})-(\ref{eq:53}),\\ $j=j+1$ };
    
    \node [cloud,below of=argmax,minimum width=2cm] (end) {End};
    \path [line] (start) -- (readimg);
    \path [line] (readimg) -- (initialize);
    \path [line] (initialize) -- (converged);
    \draw[->] (jk) |- node[below]{Yes} (VBE);
    \path [line] (VBE) -- (forwardbackward);
    \draw[->] (jk) -| node[above]{No} (converged);
    \draw[->] (converged) -- node {No}(initial);
    \draw[->] (initial) -- (jk);
    \path [line] (converged) -- node {Yes}(compute);
    \path [line] (compute) -- (argmax);
    \path [line] (argmax) -- (end);
    \path [line] (forwardbackward) -- (VBM);
    \draw[->] (VBM)-- ++(6cm,0) |-  (jk);
    
    \begin{scope}
    \node[rectangle,draw,minimum width=1cm,double=blue, double distance =1pt][fit = (compute)(argmax),color=blue!30] (SA) {};
    \node [above left] at (compute.north  east) { };
    \end{scope}
    
    \begin{scope}
    \node[rectangle,draw,minimum width=1cm][fit = (VBE)(forwardbackward)(VBM),color=red,double=red, double distance =1pt] (SA) {};
    \node [above left] at (forwardbackward.north  east) {};
    \end{scope}
    
     \node [rectangle, draw=white, text width=3cm, text centered,text=red, rounded corners, minimum height=.5cm,  above of= VBE,node distance=.9cm] (initial) {Done for all $j\in[1,K]$};
    
\end{tikzpicture}
\caption{Flowchart of the IGDTM}
\label{fig:5}       
\end{figure*}
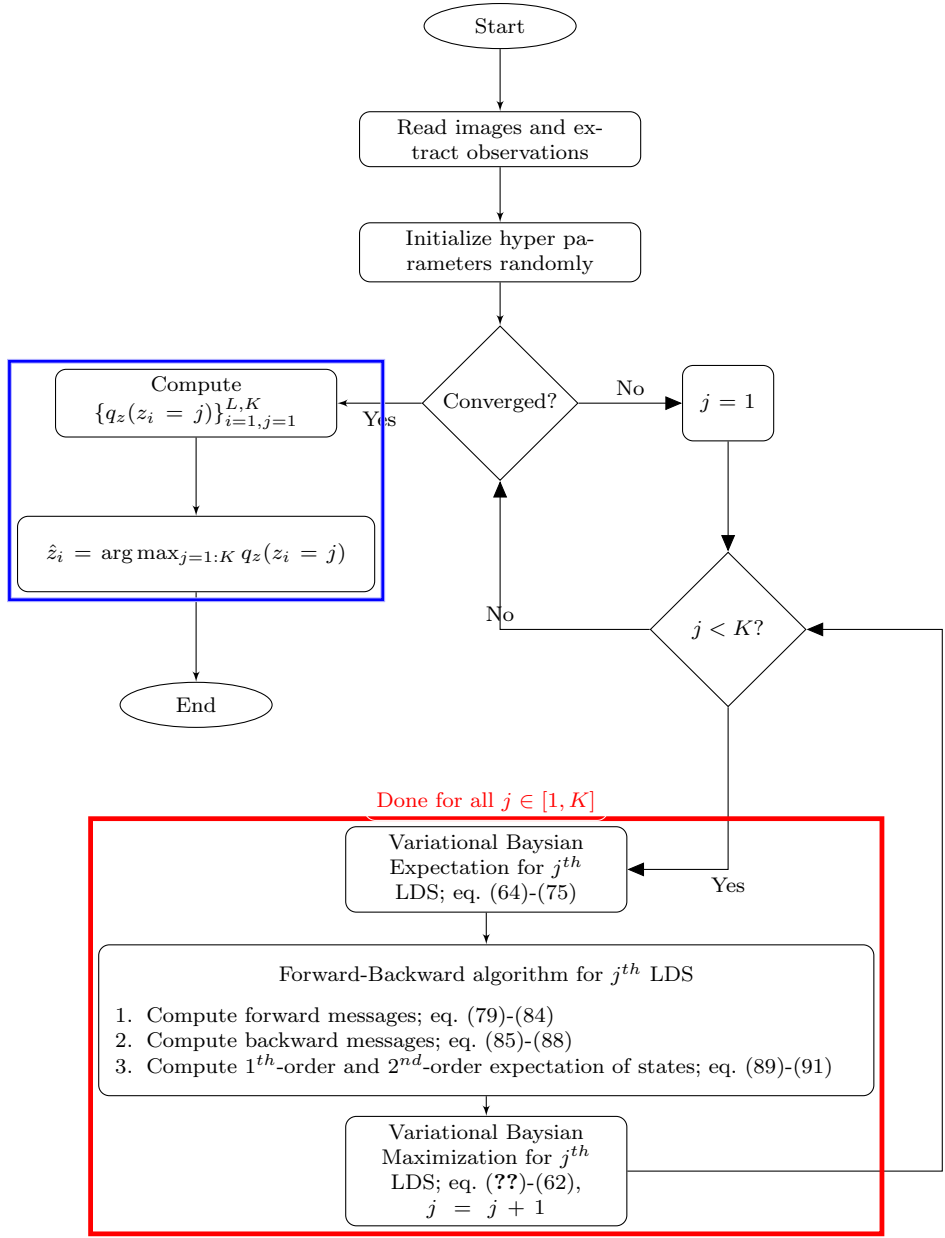

The flowchart of the IGDTM, with respect to the derived equations in previous sections, is demonstrated in figure \ref{fig:5}. Firstly, the hyper-parameters are initialized randomly. Then, while the convergence condition is not satisfied, for all of the LDSs, the Bayesian variational EM algorithm, (which composed of VBE-step, RTSS algorithm, and VBM-step), is performed.  The VBEM steps are surrounded by the red rectangle. Finally, the label field is computed by the steps are surrounded by the blue rectangle.

\begin{figure*}
\tikzset{main/.style={circle, minimum size = .2cm, thick, draw =black!80, node distance = 3mm},
         connect/.style={-latex, thick},
         box/.style={rectangle, draw=black}
}
\begin{tikzpicture}[x=2mm,
    y=2mm,chessboard/.pic={
    \tikzstyle{dot1} = [fill=white, draw];
    \tikzstyle{dot2} = [fill=black, draw];
    \tikzstyle{dot3} = [fill=white, draw];
    \tikzstyle{dot4} = [fill=black, draw];
    \tikzstyle{dot5} = [fill=black, draw];
    \foreach \x in {1, ..., 10} {
      \foreach \y in {1, ..., 10} {
        \pgfmathsetmacro\sy{random(1, 5)}%
        \path[dot\sy] (\x, \y) rectangle(10,10);
      }}
    }, lds/.pic={j=#1,
      \node[box,draw=white] (-Latent) {};
      \node[main,minimum size=.2cm, right=of -Latent] (-L1) {$x_1^{(\j)}$};
      \node[main,minimum size=.2cm] (-L2) [right=of -L1] {$x_2^{(\j)}$};
      \node[main,minimum size=.2cm] (-Lt) [right=5mm of -L2] {$x_T^{(\j)}$};
      \node[box,draw=white!100] (-Observed) [below=of -Latent] {};
      \node[main,fill=black,text=white,minimum size=.2cm] (-O1) [right=of -Observed,below=of -L1] {$y_1^{(\j)}$};
      \node[main,fill=black,text=white,minimum size=.2cm] (-O2) [right=of -O1,below=of -L2] {$y_2^{(\j)}$};
      \node[main,fill=black,text=white,minimum size=.2cm] (-Ot) [right=of -O2,below=of -Lt] {$y_T^{(\j)}$};
      \draw (-L1.east) edge [connect] (-L2);
      \node at (14,0) {$\dots$};
      \path (-L1.south) edge [connect] (-O1);
      \path (-L2.south) edge [connect] (-O2);
      \path (-Lt.south) edge [connect] (-Ot);
           },my text/.style={rounded corners=2pt, text width=10mm, font=\sffamily, line width=.5pt, align=left},
      my arrow/.style={rounded corners=2pt, draw=red!15, line width=1.5mm, -{Triangle[]}},x=2mm,
    y=2mm,blackboard/.pic={
      \path[fill=black, draw] (1,1) rectangle(10,10);
    },]

    
    \pic(cb5) at (22,11){chessboard};
    \pgfmathsetmacro\j{7} 
    \def\colr{white}
    \pic(lds5) at (0,20){lds};
    
    \pic(cb4) at (22,24){chessboard};
    \pgfmathsetmacro\j{6}
    \pic(lds4) at (0,33){lds};
    
    \pic(cb3) at (22,37){chessboard};
    \pgfmathsetmacro\j{5}
    \pic(lds3) at (0,46){lds};
    
    \pic(cb2) at (22,50){chessboard};
    \pgfmathsetmacro\j{4}
    \pic(lds2) at (0,59){lds};
    
    \pic(cb1) at (22,63){chessboard};
    \pgfmathsetmacro\j{3}
    \pic(lds1) at (0,72){lds};
    
    \pic(cb7) at (22,76){chessboard};
    \pgfmathsetmacro\j{2}
    \pic(lds7) at (0,85){lds};
    
    \pic(cb8) at (22,89){chessboard};
    \pgfmathsetmacro\j{1}
    \pic(lds8) at (0,98){lds};

    \node [
        draw, rectangle,rounded corners=10,red,
        anchor=south west,
        minimum width=6.6cm, minimum height=18.2cm,
        label= IGDTM\textquotesingle LDSs initialization,
      ] at (0,10) {};
    \path [my arrow] (33,55) -- (37,55);


    \begin{scope}
    \clip[postaction={fill=black, draw=black, line width=6mm}] (58,12) rectangle +(9,9);
    \clip[postaction={fill=white, line width=6mm}] (61,15) rectangle +(3,3);
    \end{scope}
    \filldraw[fill=black] (62,16) rectangle (63,17);
    \pgfmathsetmacro\j{7} 
    \pic(lds5) at (36,20){lds};
    
    \begin{scope}
    \clip[postaction={fill=white, draw=black, line width=6mm}] (58,25) rectangle +(9,9);
    \clip[postaction={fill=black, line width=6mm}] (61,28) rectangle +(3,3);
    \end{scope}
    \pgfmathsetmacro\j{6}
    \pic(lds4) at (36,33){lds};
    
    \pic(cb3) at (57,37){blackboard};
    \pgfmathsetmacro\j{5}
    \pic(lds3) at (36,46){lds};
    
    \pic(cb2) at (57,50){blackboard};
    \pgfmathsetmacro\j{4}
    \pic(lds2) at (36,59){lds};
    
    \pic(cb1) at (57,63){blackboard};
    \pgfmathsetmacro\j{3}
    \pic(lds1) at (36,72){lds};
    
    \begin{scope}
    \clip[postaction={fill=black, draw=white, line width=6mm}] (58,77) rectangle +(9,9);
    \clip[postaction={fill=black, line width=6mm}] (61,80) rectangle +(3,3);
    \end{scope}
    \draw (58,77) rectangle (67,86);
    \pgfmathsetmacro\j{2}
    \pic(lds7) at (36,85){lds};
    
    \pic(cb8) at (57,89){blackboard};
    \pgfmathsetmacro\j{1}
    \pic(lds8) at (36,98){lds};
    
    \node [
        draw, rectangle,rounded corners=10,green,
        anchor=south west,
        minimum width=6.2cm, minimum height=18.2cm,
        label= IGDTM segmentation result,
      ] at (37,10) {};
      
      \node (image) at (20,3) {\includegraphics[width=2cm]{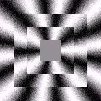}} ;
      \node[below right] at (15,-2) {Original image};
      \node (result) at (50,3) {\includegraphics[width=2cm]{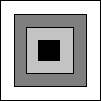}};
      \node[below right] at (40,-2) {IGDTM segmentation result};
      
      \path [my arrow] (32,3) -- (38,3);
  \end{tikzpicture}
  \caption{IGDTM segmentation for $K=7$, Red plate: The parameters of $K$ LDSs and their corresponding label field is initialized randomly, Greet plate: after performing IGDTM the number of textures is determined automatically, The non-black label fields illustrate their corresponding LDSs which are considered for segmentation.}
  \label{fig:5_1}
\end{figure*}

Figure \ref{fig:5_1} illustrates the process of IGDTM segmentation for a frame video contains four segments for the initial value of $K=7$. 

\section{Experimental Result}
\label{sec:9}
\begin{figure}
  \centering
  \subfigANDtitle{\includegraphics[width=.3\textwidth]{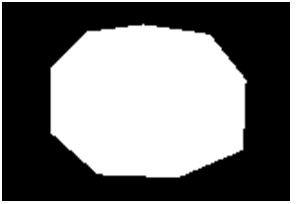} \\ (a)} \hfill
  \subfigANDtitle{\includegraphics[width=.3\textwidth]{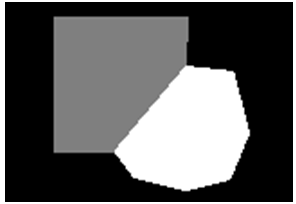} \\ (b)} \hfill
  \hfill\mbox{}
  \caption{Initial contours of the DTM\cite{chan2008modeling}, (a) Initial contour of synthdb\_2K, (b) Initial contour of synthdb\_3K}
  \label{fig:6}
\end{figure}

\begin{figure}
\centering
  \subfigANDtitle{\includegraphics[width=.3\textwidth]{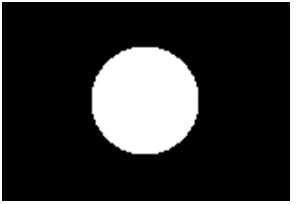} \\ (a)} \hfill
  \subfigANDtitle{\includegraphics[width=.3\textwidth]{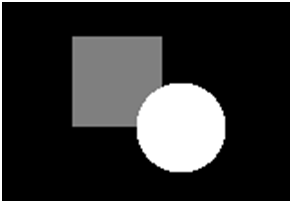} \\ (b)} \hfill
  \hfill\mbox{}
  \caption{Ground truth of UCSD\_Synthdb dataset, (a) Ground truth of synthdb\_2K, (b) Ground truth of synthdb\_3K}
  \label{fig:7}
\end{figure}

Here the results of the IGDTM, the DTM \cite{chan2008modeling} and the LDT \cite{chan2009layered} and SAW \cite{gonccalves2013dynamic} are compared. We evaluated the proposed method on three types of textures: microscopic, macroscopic and objects. For this goal, we used UCSD\_Synthdb \cite{chan2008modeling},  Dyntex \cite{peteri2010dyntex} and UCSD\_Pedestrian \cite{chan2008modeling} datasets. UCSD\_Synthdb contains synthetic video sequences consist of microscopic and macroscopic textures such as smoke, fire, sea water, vegetation, escalator etc. Dyntex is a comprehensive database of dynamic textures providing a large and diverse database of high-quality dynamic textures, which have been deinterlaced with a spatiotemporal median filter. UCSD\_Pedestrian contains video of pedestrians on UCSD walkways, from two viewpoints taken with a stationary camera. Table \ref{tab:3} illustrates the properties of the datasets in detail. As before mentioned, the DTM uses an initial contour and LDT uses the result DTM as the initial partition. Figure \ref{fig:6} illustrates the initial contour of DTM for UCSD\_Synthdb segmentation. Moreover, the ground truth of the UCSD\_Synthdb for videos with 2 and 3 segments are shown in figure \ref{fig:7}.

\begin{table}
\centering
\caption{Please write your table caption here}
\label{tab:3}       
\begin{tabular}{lll}
\hline\noalign{\smallskip}
Dataset & Resolution & Textures  \\
\noalign{\smallskip}\hline\noalign{\smallskip}
UCSD\_Synthdb & 160$\times$110 & Multiple co-occurring textures in a single video \\
Dyntex & 352$\times$288 & Sea, Grass, Trees, Vegetation, Calm water, \\
&& Fountains, Smoke, Traffic, etc.\\
UCSD\_Pedestrian & 238$\times$158 & Pedestrians on UCSD walkways \\
\noalign{\smallskip}\hline
\end{tabular}
\end{table}

\begin{figure}
  \centering
  \subfigANDtitle{\includegraphics[width=.22\textwidth]{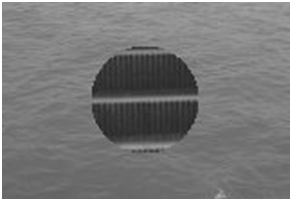}\\} \hfill
  \subfigANDtitle{\includegraphics[width=.22\textwidth]{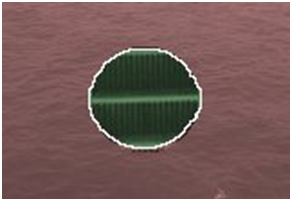}\\r=0.9728} \hfill
  \subfigANDtitle{\includegraphics[width=.22\textwidth]{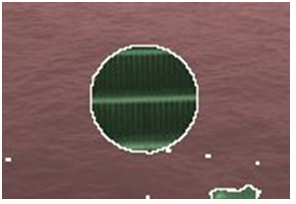}\\r=0.9839} \hfill
  \subfigANDtitle{\includegraphics[width=.22\textwidth]{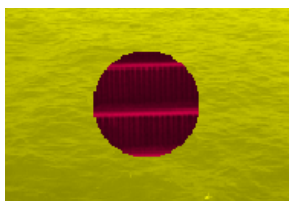}\\r=1\\ IGDTM\_Seg\_no=2} \hfill

  \subfigANDtitle{\includegraphics[width=.22\textwidth]{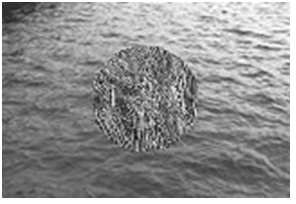}\\ \\} \hfill
  \subfigANDtitle{\includegraphics[width=.22\textwidth]{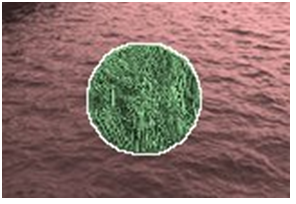}\\r=0.9618\\} \hfill
  \subfigANDtitle{\includegraphics[width=.22\textwidth]{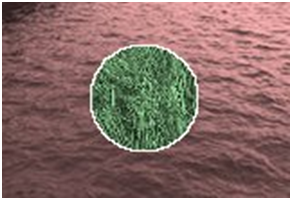}\\r=0.9990\\} \hfill
  \subfigANDtitle{\includegraphics[width=.22\textwidth]{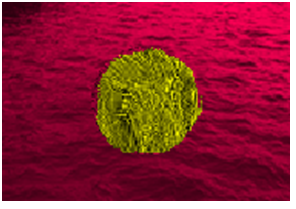}\\r=0.9838\\ IGDTM\_Seg\_no=2} \hfill
  
  \subfigANDtitle{\includegraphics[width=.22\textwidth]{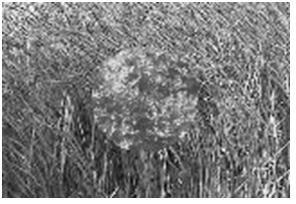}\\ \\} \hfill
  \subfigANDtitle{\includegraphics[width=.22\textwidth]{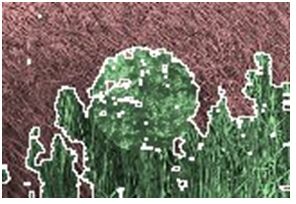}\\r=0.5467\\} \hfill
  \subfigANDtitle{\includegraphics[width=.22\textwidth]{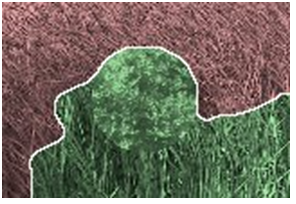}\\r=0.5534\\} \hfill
  \subfigANDtitle{\includegraphics[width=.22\textwidth]{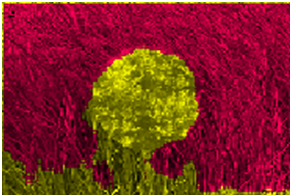}\\r=0.8591\\ IGDTM\_Seg\_no=2} \hfill
  
  \subfigANDtitle{\includegraphics[width=.22\textwidth]{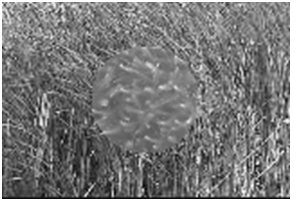}\\ \\} \hfill
  \subfigANDtitle{\includegraphics[width=.22\textwidth]{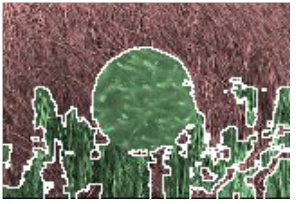}\\r=0.6080\\} \hfill
  \subfigANDtitle{\includegraphics[width=.22\textwidth]{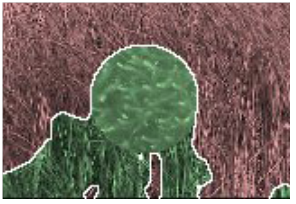}\\r=0.6885\\} \hfill
  \subfigANDtitle{\includegraphics[width=.22\textwidth]{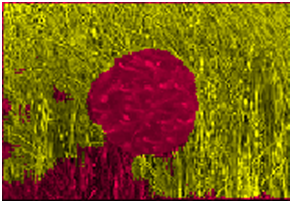}\\r=0.7288\\ IGDTM\_Seg\_no=2} \hfill
  
  \subfigANDtitle{\includegraphics[width=.22\textwidth]{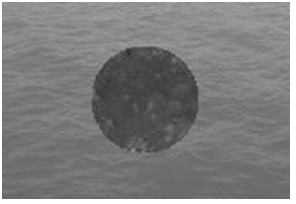}\\ \\} \hfill
  \subfigANDtitle{\includegraphics[width=.22\textwidth]{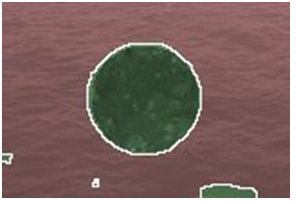}\\r=0.9436\\} \hfill
  \subfigANDtitle{\includegraphics[width=.22\textwidth]{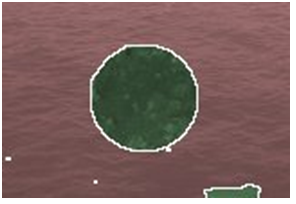}\\r=0.9800\\} \hfill
  \subfigANDtitle{\includegraphics[width=.22\textwidth]{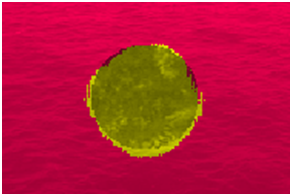}\\r=0.9810\\ IGDTM\_Seg\_no=2} \hfill
  
  \hfill\mbox{}
  \caption{Results on the UCSD\_synthdb database (synthdb\_2K), left to right: a frame of the original video, the segmentation result of the DTM \cite{chan2008modeling}, the segmentation result of the LDT \cite{chan2009layered}, the segmentation result of the IGDTM; The r-index and IGDTM\_Seg\_no of the segmentation is given below the images.}
  \label{fig:8}
\end{figure}

\begin{figure}
  \centering
  \subfigANDtitle{\includegraphics[width=.22\textwidth]{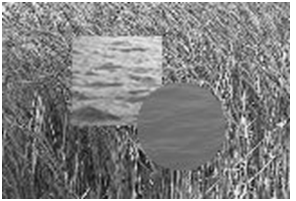}\\} \hfill
  \subfigANDtitle{\includegraphics[width=.22\textwidth]{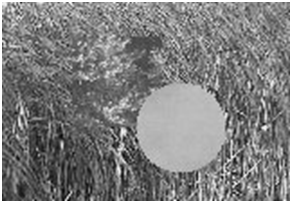}\\r=0.6588} \hfill
  \subfigANDtitle{\includegraphics[width=.22\textwidth]{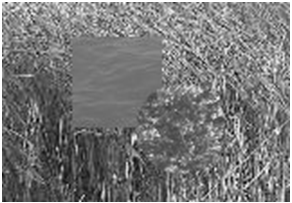}\\r=0.6820} \hfill
  \subfigANDtitle{\includegraphics[width=.22\textwidth]{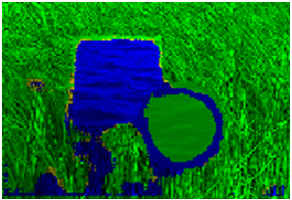}\\r=0.7483\\ IGDTM\_Seg\_no=3} \hfill

  \subfigANDtitle{\includegraphics[width=.22\textwidth]{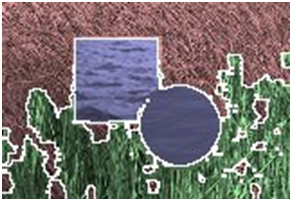}\\ \\} \hfill
  \subfigANDtitle{\includegraphics[width=.22\textwidth]{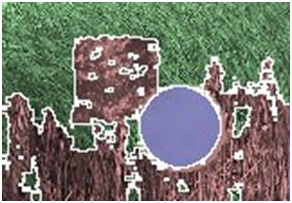}\\r=0.5951\\} \hfill
  \subfigANDtitle{\includegraphics[width=.22\textwidth]{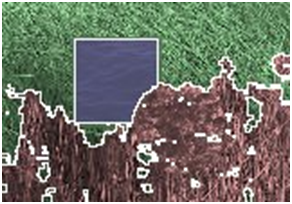}\\r=0.6163\\} \hfill
  \subfigANDtitle{\includegraphics[width=.22\textwidth]{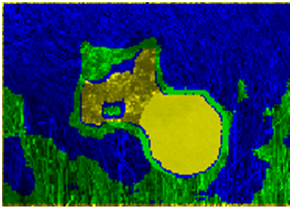}\\r=0.7893\\ IGDTM\_Seg\_no=3} \hfill
  
  \subfigANDtitle{\includegraphics[width=.22\textwidth]{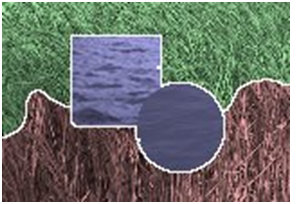}\\ \\} \hfill
  \subfigANDtitle{\includegraphics[width=.22\textwidth]{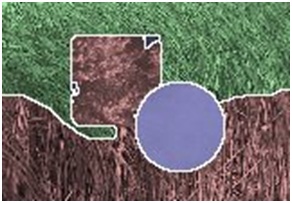}\\r=0.6101\\} \hfill
  \subfigANDtitle{\includegraphics[width=.22\textwidth]{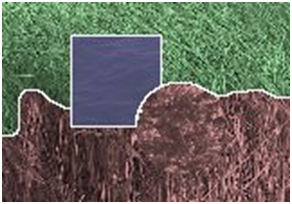}\\r=0.6272\\} \hfill
  \subfigANDtitle{\includegraphics[width=.22\textwidth]{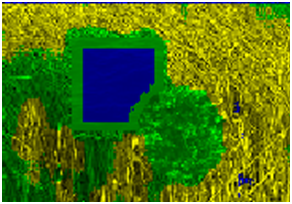}\\r=0.7276\\ IGDTM\_Seg\_no=3} \hfill
  
  \hfill\mbox{}
  \caption{Results on the UCSD\_synthdb database (synthdb\_3K), left to right: a frame of the original video, the segmentation result of the DytexMicIC \cite{chan2008modeling}, the segmentation result of the LDT \cite{chan2009layered}, the segmentation result of the IGDTM, The r-index and IGDTM\_Seg\_no of the segmentation is given below the images.}
  \label{fig:9}
\end{figure}

Figure \ref{fig:8} illustrates the results of IGDTM, DTM \cite{chan2008modeling} and LDT \cite{chan2009layered} on the UCSD\_Synthdb dataset for two dynamic textures. Rand index (r-index) of the segmentation results are given below the images. The dynamic texture numbers, which are estimated by IGDTM is defined by ``IGDTM\_Seg\_no'' tag, are given below the IGDTM results. 
Figure \ref{fig:9} illustrates the results of IGDTM, DTM and LDT on the UCSD\_Synthdb dataset for three dynamic textures. Similar to the previous figure, the Rand index of the segmentation results are given below the images. 

\begin{figure}
  \centering
  \subfigANDtitle{\includegraphics[width=.35\textwidth]{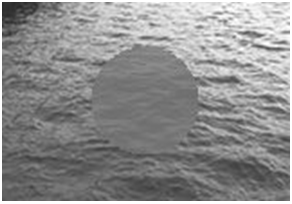}\\ \\(a)} \hfill
  \subfigANDtitle{\includegraphics[width=.35\textwidth]{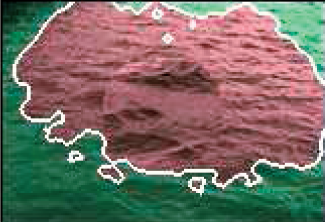}\\r=0.5069\\(b)} \hfill
  \subfigANDtitle{\includegraphics[width=.35\textwidth]{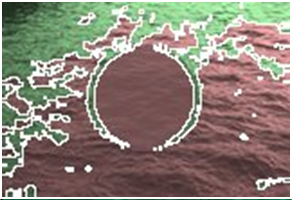}\\r=0.5060\\(c)} \hfill
  
  \subfigANDtitle{\includegraphics[width=.35\textwidth]{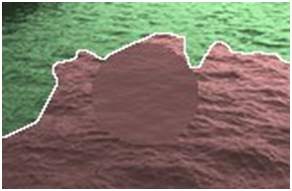}\\r=0.5001\\(d)} \hfill  
  \subfigANDtitle{\includegraphics[width=.35\textwidth]{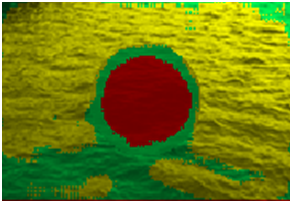}\\r=0.6702\\ IGDTM\_Seg\_no=3\\(e)} \hfill
  \hfill\mbox{}
  \caption{Segmentation results for a video sequence of UCSD\_synthdb\_2K, (a) a video frame, segmentation with: (b) Ising initialized using DTM \cite{ghoreyshi2007segmenting}, (c) DTM, (d) LDT, (e) IGDTM; The r-index and IGDTM\_Seg\_no of the segmentation is given below the images.}
  \label{fig:10}
\end{figure}

As mentioned before, the segmentation process in DTM and LDT is done subjectively using expert knowledge which is a drastic limitation for systematic database approaches. The IGDTM eliminates the mentioned restriction. Therefore, the method finds the optimum textures number automatically. Figure \ref{fig:10} and 11 illustrates two results which IGDTM over segmented the video sequence. These videos are segmented by one more texture number. 

Table \ref{tab:4} shows the result of segmentation for IGDTM, LDT and DTM on Synthdb dataset quantitatively based on average R-index. 
 
 \begin{figure}
  \centering
  \subfigANDtitle{\includegraphics[width=.35\textwidth]{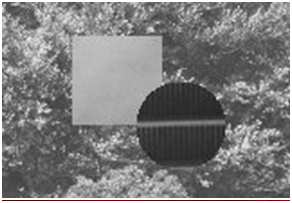}\\} \hfill
  \subfigANDtitle{\includegraphics[width=.35\textwidth]{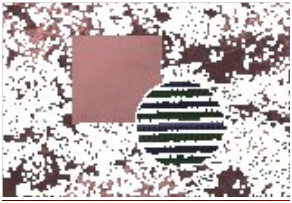}\\r=0.5373} \hfill
  \subfigANDtitle{\includegraphics[width=.35\textwidth]{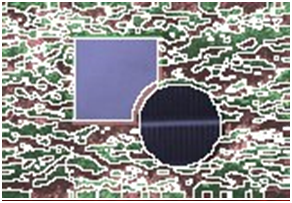}\\r=0.6549} \hfill
  
  \subfigANDtitle{\includegraphics[width=.35\textwidth]{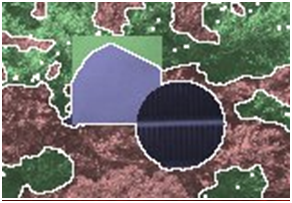}\\r=0.6563\\ } \hfill
  \subfigANDtitle{\includegraphics[width=.35\textwidth]{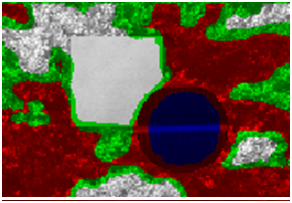}\\r=0.7672\\ IGDTM\_Seg\_no=4} \hfill
  \hfill\mbox{}
  \caption{Segmentation results for a video sequence of UCSD\_synthdb\_3K, (a) a video frame of UCSD\_Synthdb\_3K, segmentation with: (b) GPCA \cite{vidal2005optical}, (c) DTM, (d) LDT, (e) IGDTM; The r-index and IGDTM\_Seg\_no of the segmentation is given below the images.}
  \label{fig:11}
\end{figure}

\begin{figure}
  \centering
  \subfigANDtitle{\includegraphics[width=.22\textwidth]{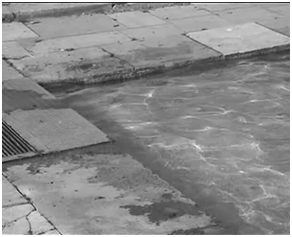}\\} \hfill
  \subfigANDtitle{\includegraphics[width=.22\textwidth]{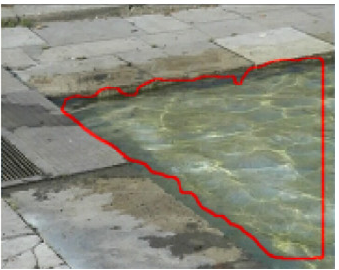}\\} \hfill
  \subfigANDtitle{\includegraphics[width=.22\textwidth]{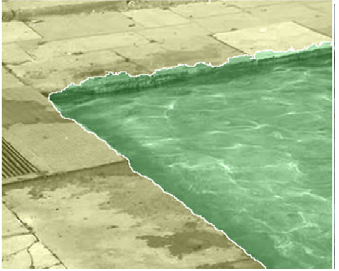}\\} \hfill
  \subfigANDtitle{\includegraphics[width=.22\textwidth]{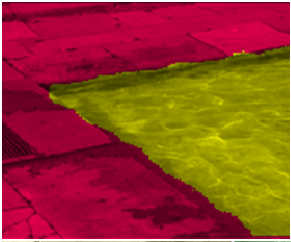}\\ } \hfill
  \subfigANDtitle{IGDTM\_Seg\_no=2 } \hfill
  
  \subfigANDtitle{\includegraphics[width=.22\textwidth]{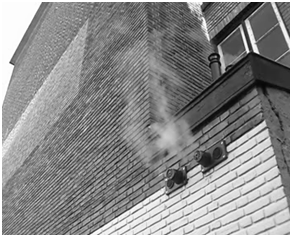}\\} \hfill
  \subfigANDtitle{\includegraphics[width=.22\textwidth]{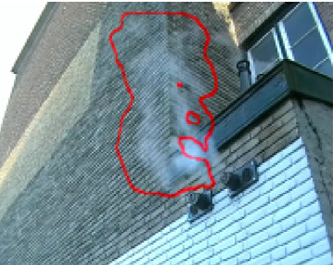}\\} \hfill
  \subfigANDtitle{\includegraphics[width=.22\textwidth]{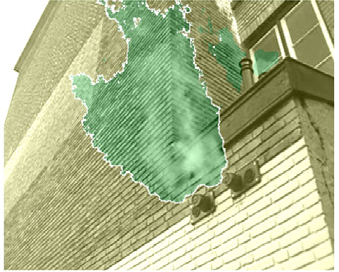}\\} \hfill
  \subfigANDtitle{\includegraphics[width=.22\textwidth]{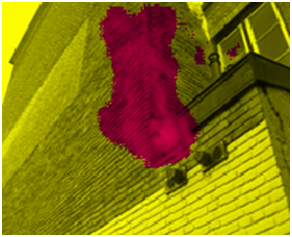}\\ } \hfill
  \subfigANDtitle{IGDTM\_Seg\_no=2 } \hfill
  
  \subfigANDtitle{\includegraphics[width=.22\textwidth]{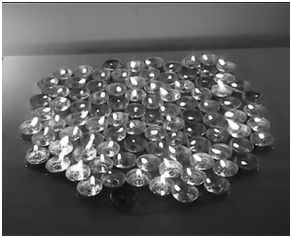}\\} \hfill
  \subfigANDtitle{\includegraphics[width=.22\textwidth]{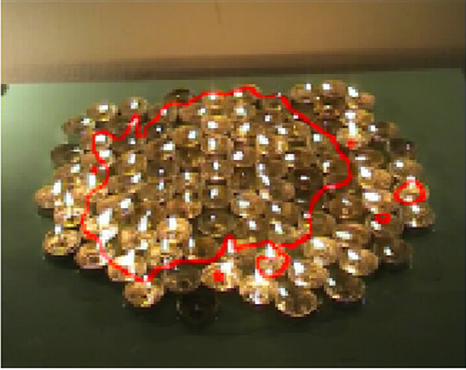}\\} \hfill
  \subfigANDtitle{\includegraphics[width=.22\textwidth]{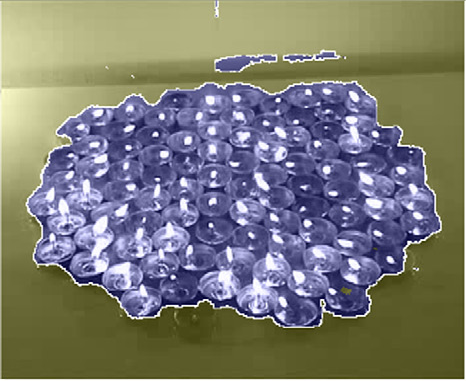}\\} \hfill
  \subfigANDtitle{\includegraphics[width=.22\textwidth]{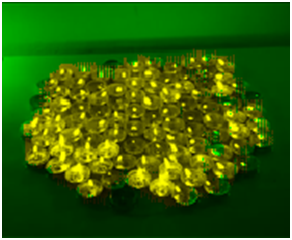}\\ } \hfill
  \subfigANDtitle{IGDTM\_Seg\_no=2 } \hfill
  
  \hfill\mbox{}
  \caption{Segmentation result for Dyntex datasets, from left to right: a frame of the video, segmentation with: Proposed method in \cite{fazekas2009dynamic}, SAW \cite{gonccalves2013dynamic},  IGDTM, DT\textquotesingle s number estimated by IGDTM.}
  \label{fig:12}
\end{figure}

Figure \ref{fig:12} illustrates the segmentation results of IGDTM in comparison with the proposed method in \cite{fazekas2009dynamic} and \cite{gonccalves2013dynamic}. As the results show, despite the elimination of the initial contour and expert knowledge, IGDTM overcomes the segmentation problem properly. Figure 13 illustrates the segmentation results of IGDTM on some of the video sequences of Dyntex dataset qualitatively.
\begin{table}
\caption{Rand index comparison for some approaches}
\label{tab:4}       
\centering
\begin{tabular}{lll}
\hline\noalign{\smallskip}
Method & Synthdb\_2K & Synthdb\_3K  \\
\noalign{\smallskip}\hline\noalign{\smallskip}
LDT\cite{chan2009layered} & 0.942 & 0.921 \\
DTM\cite{chan2008modeling} & 0.892 & 0.841 \\
IGDTM &  &  \\
\noalign{\smallskip}\hline
\end{tabular}
\end{table}

\begin{figure}
  \centering
  \subfigANDtitle{\includegraphics[width=.22\textwidth]{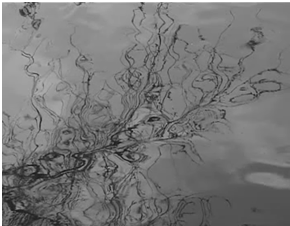}\\} \hfill
  \subfigANDtitle{\includegraphics[width=.22\textwidth]{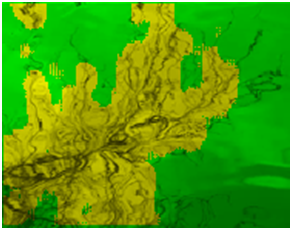}\\IGDTM\_Seg\_no=2} \hfill
  
  \subfigANDtitle{\includegraphics[width=.22\textwidth]{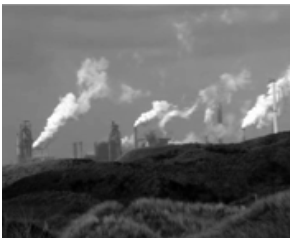}\\} \hfill
  \subfigANDtitle{\includegraphics[width=.22\textwidth]{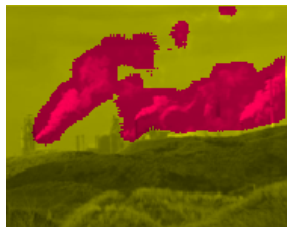}\\IGDTM\_Seg\_no=2} \hfill
  
  \subfigANDtitle{\includegraphics[width=.22\textwidth]{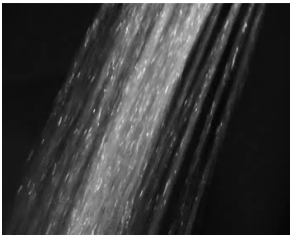}\\} \hfill
  \subfigANDtitle{\includegraphics[width=.22\textwidth]{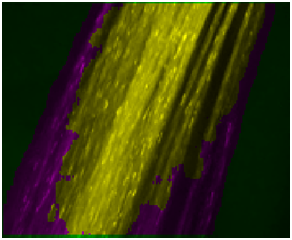}\\IGDTM\_Seg\_no=3} \hfill
  
  \subfigANDtitle{\includegraphics[width=.22\textwidth]{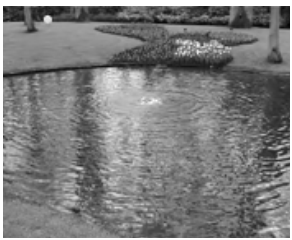}\\ } \hfill
  \subfigANDtitle{\includegraphics[width=.22\textwidth]{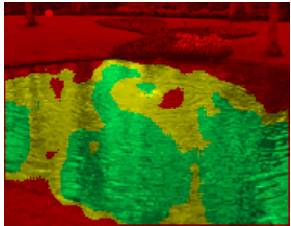}\\IGDTM\_Seg\_no=3 } \hfill
  
  \subfigANDtitle{\includegraphics[width=.22\textwidth]{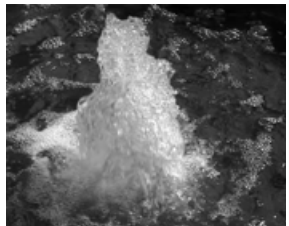}\\ } \hfill
  \subfigANDtitle{\includegraphics[width=.22\textwidth]{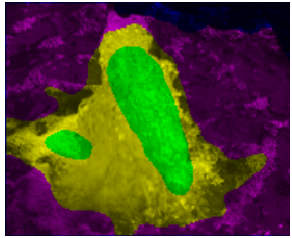}\\IGDTM\_Seg\_no=4 } \hfill

  \hfill\mbox{}
  \caption{Segmentation results of IGDTM for Dyntex datasets, left to right: a frame of video sequence,  segmentation result of IGDTM, DT\textquotesingle s number estimated by IGDTM is shown below of the results}
  \label{fig:13}
\end{figure}

\begin{figure}
  \centering
  \subfigANDtitle{\includegraphics[width=.22\textwidth]{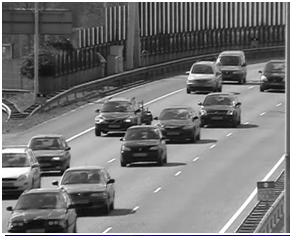}\\} \hfill
  \subfigANDtitle{\includegraphics[width=.22\textwidth]{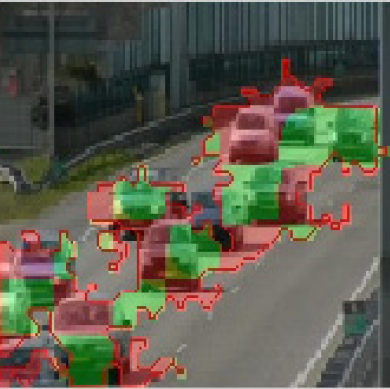}\\} \hfill
  \subfigANDtitle{\includegraphics[width=.22\textwidth]{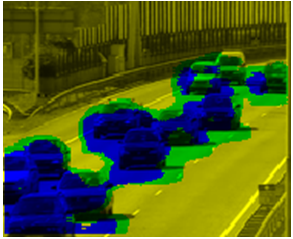}\\} \hfill
  \subfigANDtitle{\includegraphics[width=.22\textwidth]{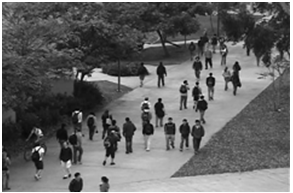}\\IGDTM\_Seg\_no=2} \hfill
  \subfigANDtitle{\includegraphics[width=.22\textwidth]{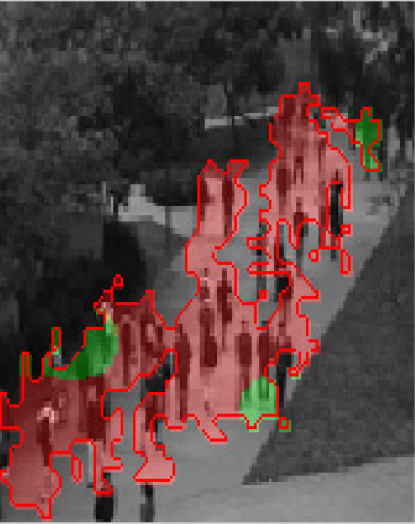}\\IGDTM\_Seg\_no=2} \hfill
  \subfigANDtitle{\includegraphics[width=.22\textwidth]{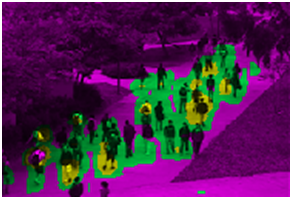}\\IGDTM\_Seg\_no=3} \hfill
  
  \hfill\mbox{}
  \caption{Segmentation result for Dyntex datasets, from left to right: a frame of the video, segmentation with: the proposed method in \cite{soygaonkar2015dynamic}, IGDTM, DT\textquotesingle s number estimated by IGDTM is shown below of the results}
  \label{fig:14}
\end{figure}
\begin{figure}
  \centering
  \subfigANDtitle{\includegraphics[width=.22\textwidth]{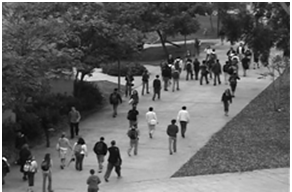}\\} \hfill
  \subfigANDtitle{\includegraphics[width=.22\textwidth]{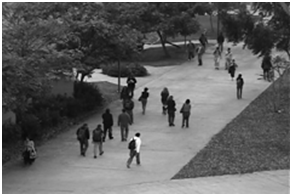}\\} \hfill
  \subfigANDtitle{\includegraphics[width=.22\textwidth]{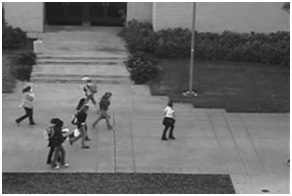}\\} \hfill
  \subfigANDtitle{\includegraphics[width=.22\textwidth]{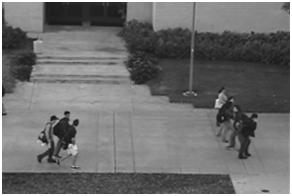}\\} \hfill
  \subfigANDtitle{\includegraphics[width=.22\textwidth]{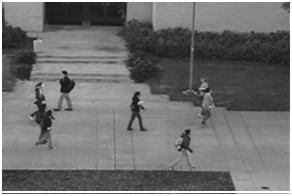}\\} \hfill
  
  \subfigANDtitle{\includegraphics[width=.22\textwidth]{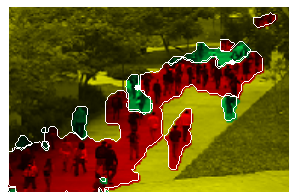}\\} \hfill
  \subfigANDtitle{\includegraphics[width=.22\textwidth]{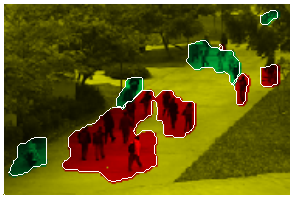}\\} \hfill
  \subfigANDtitle{\includegraphics[width=.22\textwidth]{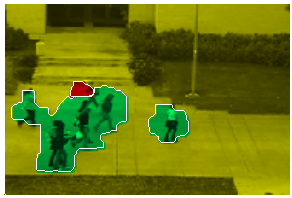}\\} \hfill
  \subfigANDtitle{\includegraphics[width=.22\textwidth]{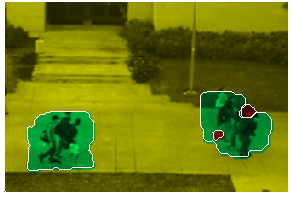}\\} \hfill
  \subfigANDtitle{\includegraphics[width=.22\textwidth]{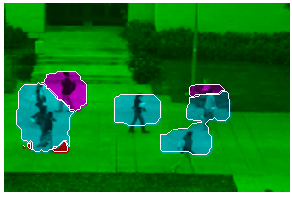}\\} \hfill
  
  \subfigANDtitle{\includegraphics[width=.22\textwidth]{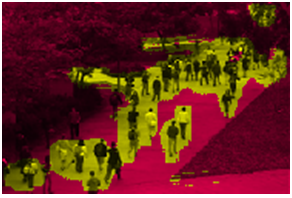}\\IGDTM\_Seg\_no=2} \hfill
  \subfigANDtitle{\includegraphics[width=.22\textwidth]{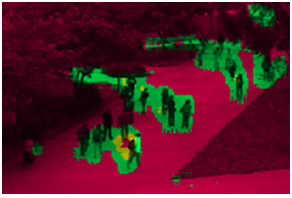}\\IGDTM\_Seg\_no=3} \hfill
  \subfigANDtitle{\includegraphics[width=.22\textwidth]{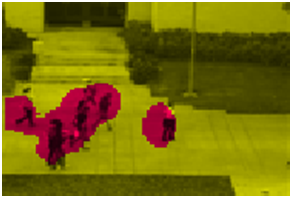}\\IGDTM\_Seg\_no=2} \hfill
  \subfigANDtitle{\includegraphics[width=.22\textwidth]{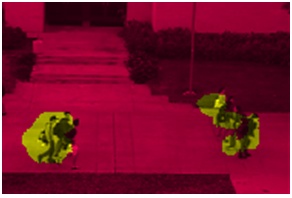}\\IGDTM\_Seg\_no=2} \hfill
  \subfigANDtitle{\includegraphics[width=.22\textwidth]{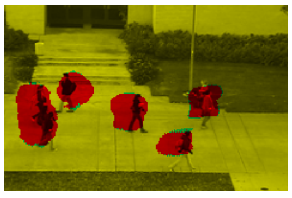}\\IGDTM\_Seg\_no=3} \hfill
  
  \hfill\mbox{}
  \caption{Segmentation result for UCSD\_Pedestrian datasets, from top to down: a frame of the video, ground truth, segmentation with IGDTM, DT\textquotesingle s number estimated by IGDTM.}
  \label{fig:15}
\end{figure}

Ultimately, we present segmentation results of some object-based-dynamic-textures. Figure \ref{fig:14} shows the segmentation results of two object-based dynamic textures from the Dyntex dataset. The IGDTM segmentation results is compared with the proposed method in \cite{soygaonkar2015dynamic}. As results show, the IGDTM produces the segmentation results with smoother boundaries. Moreover, figure \ref{fig:15} illustrates the segmentation results for some of the video sequences from UCSD\_Pedestrian dataset in compared with the ground truth available in the dataset. We set the value of $K$ to $20$.

\section{Conclusions}
\label{sec:10}
The aim of this paper was to propose a proper dynamic texture segmentation approach which eliminates the expert knowledge about the number of the dynamic textures and initial partitioning in the previous methods. For this goal, we introduced a novel fully Bayesian non-parametric formulation of generative dynamic texture models for robust unsupervised dynamic texture segmentation. By deriving a suitable posterior distribution over the number of textures, the proposed model resolved the problem of determining the proper number of texture segmentation automatically. Because of a better performance of Variational Bayesian approximation methods compared with Monte Carlo techniques, we used the Variational Bayesian EM for inference. In order to compute the 1st order and 2nd order sufficient statistics of the hidden states, Variational Bayesian Rauch-Tung-Striebel smoother (RTSS) is applied. Finally, a comprehensive comparison of the proposed method with the state of the art methods, for the three dynamic texture categories (i.e. microscopic, macroscopic and object based dynamic texture), was presented. 

\appendix
\begin{appendices}
\section{Variational Bayesian M-step }
\label{sec:appA}
As mentioned before, the Variational posterior is expected to be the same distributed from $p(\Psi\mid\Xi,y)$ \cite{Bishop2006}. Since we defined the conjugate exponential priors, the Variational posterior will be distributed from the exponential family. It is shown that the best distribution for $q_j^*$, for each of the factors $q_j$ (in terms of the distribution minimizing the KL measurement as described before, equations (\ref{eq:21}),(\ref{eq:22})) is expressed as 

\begin{equation}\label{eq:93}
    \ln q_\Psi^*(\Psi_j\mid y,\Xi)=E_{\setminus j}[\ln p\left(\Psi,\Xi,y\right)]+const,
\end{equation}

where $E_{\setminus j}[.]$ is the expectation taken over all variables not in the partition \cite{Bishop2006}. In this appendix, the distribution of $q_j^*$ , for each of factors are computed. 
\subsection{Estimation of the distribution $q(z_{i^\prime})$?}
\label{sec:12}
According to what explained before, for estimating $q(z_{i^\prime})$  we can write:

\begin{multline}\label{eq:94}
    P\left(z_{i}\mid x_{1:T},z_{i\neq i},\pi_{1:K},\alpha,\delta_{1:K},S_{1:K},Q_{1:K},\mu_{1:K},\Sigma_{1:K},R_{1:K},\Xi\right)\\
     =\prod_{j=1}^K N(x_1^{(j)}\mid\delta_j,S_j^{-1})^{\mathbb{I}(z_{i}=j)}\prod_{t=2}^T N(x_t^{(j)}\mid Ax_t^{(j)},Q_j^{-1})^{\mathbb{I}(z_{i}=j)}\\
     \times\prod_{t=1}^T\prod_{j=1}^K\left(\pi_j\left(\nu\right)N\left(y_{i^\prime t}\mid C^{(j)}x_t^{(j)},R_j^{-1}\right)N\left(y_{i t}\mid\mu_j,\Sigma_j^{-1}\right)\right)^{\mathbb{I}(z=j)}
\end{multline}

It is easy to show that by replacing (\ref{eq:94}) into (\ref{eq:93}) and doing some calculations, the distribution of $q(z_{i})$ is obtained via (\ref{eq:40}). In which, by defining $\bar{p}_{i,j}=E[z_{i}=j]$, we can derive $q\left(z_i>j\right)=\sum_{k=j+1}^K\bar{p}_{i,k}$.

\subsection{Estimation of the distribution $q(\nu_{j})$}
\label{sec:13}
The derivation of $q(\nu_{j})$ is similar to what is seen before. We can write:

\begin{multline}\label{eq:95}
    P\left(\nu_{j}\mid v_{\backslash j},x_{1:T},Z_{1:L},\alpha,\delta_{1:K},S_{1:K},Q_{1:K},\mu_{1:K},r_{1:K},A^{1:K},C^{1:K},\Xi \right)\\
    =\prod_{i=1}^LP\left(z_i=j\mid\pi_{j}(\nu)\right)^{\mathbb{I}\left(z_i=j\right)}P\left(\nu\left(\pi_{j}\right)\mid\alpha\right).
\end{multline}

According to the equation (\ref{eq:93}), we can derive:

\begin{equation}\label{eq:96}
    q_\nu(\nu_{j})\propto\exp\left(\left(E[\alpha]+\sum_{i=1}^Lq \left(z_i=j\right) -1\right)\ln\nu_j+\sum_{i=1}^Lq\left(z_i>j\right)\ln\left(1-\nu_{j}\right)\right)
\end{equation}

It is easy to show that by simplifying (\ref{eq:96}) the distribution of $q_(\nu_{j})$ will be Beta distribution and can be rewritten as (\ref{eq:50})-(\ref{eq:52}).

\subsection{Estimation of the distribution $q\left(\delta_{j},S_{j}\right)$}
\label{sec:14} 
Similar to the previous sections, we can write
\begin{equation}\label{eq:97}
    P(\delta_{j},S_{j}\mid x_{1:T},z_{1:L},\pi_{1:K},\alpha,\delta_{\backslash j},S_{\backslash j},Q_{1:K},\mu_{1:K},r_{1:K},A^{1:K},C^{1:K})=    N\left(x_1^{(j)}\mid\delta_{j},S_{j}^{-1}\right)NW\left(\delta_{j},S_{j}\mid m_3,\lambda_3,w_3,\Psi_3\right)
\end{equation}

Therefore with respect to the equation (\ref{eq:93}), we can derive:

\begin{equation}\label{eq:98}
    q\left(\delta_{j},S_{j}\right)=\exp {E_{\setminus \delta_{j},S_{j}}\begin{pmatrix}\ln N\left( x_1^{(j)}\mid\delta_{j},S_{j}^{-1} \right)
    +\ln NW\left(\delta_{j},S_{j}\mid m_3,\lambda_3,w_3,\Psi_3\right)\end{pmatrix} }
\end{equation}

According to what described before, the distribution of $q\left(\delta_{j^\prime},S_{j^\prime}\right)$ will be Normal-Wishart. By simplifying equation (\ref{eq:98}), the distribution of $q\left(\delta_{j^\prime},S_{j^\prime}\right)$ is obtained via (\ref{eq:39})\textendash(\ref{eq:46}).

\subsection{Estimation of the distribution $q\left(\mu_j,r_{j}\right)$}
\label{sec:15} 

Similar to the previous sections, we can derive:
\begin{multline}\label{eq:99}
    P\left(\mu_j,r_{j}\mid x_{1:T},z_{1:L},\pi_{1:K},\alpha,\delta_{1:K},S_{1:K},Q_{1:K},\mu_{\backslash j},r_{\backslash j},A^{1:K},C^{1:K},\Xi\right)\\
    =\prod_{t=1}^{T}\prod_{i=1}^{L}N\left(y_{it}\mid C_i^{(j)}x_t^{(j)}+\mu_j,r_{j}^{-1}\right)^{\mathbb{I}\left(z_i=j\right)} NW\left(\mu_j,r_{j}\mid m_2,\lambda_2, w_2,\Psi_2\right).
\end{multline}

Using equations (\ref{eq:93}) and (\ref{eq:99}), the distribution of $q(\mu_j,r_{j})$ can be written:
\begin{multline}\label{eq:100}
    q\left(\mu_j,r_{j}\right)=\exp E_{\setminus \mu_j,R_{j}} \left[ \sum_{t=1}^T\sum_{i=1}^L \mathbb{I}\left(z_i=j\right)\ln{N\left(y_{it}\mid c_i^{(j)}x_t^{(j)}+\mu_j,r_j^{-1}\right)}\right.\\
    \left.    +\ln{NW\left(Q_{j}\mid  m_2,\lambda_2,w_2,\Psi_2\right)} \right]
\end{multline}

By simplifying equation (\ref{eq:100}), the distribution of   is derived as Wishart and is given by (\ref{eq:32})\textendash(\ref{eq:37}).

\subsection{Estimation of the distribution $q\left(Q_{j}\right)$}
\label{sec:16} 
In order to compute the distribution of $q\left(Q_{j}\right)$  we can write:

\begin{multline}\label{eq:101}
    P\left(Q_{j}\mid x_{1:T},z_{1:L},\pi_{1:K},\alpha,\delta_{1:K},S_{1:K},Q_{\backslash j},\mu_{1:K},r_{1:K},A^{1:K},C^{1:K},\Xi\right)\\
    =\prod_{t=2}^{T}N\left(x_{t}^{(j)}\mid A^{(j)}x_{t-1}^{(j)},Q_{j}^{-1}\right)W(Q_{j}\mid w_1,\Psi_1).
\end{multline}

So, with respect to the equation (\ref{eq:93}) and simplifying the equation (\ref{eq:101}), the distribution of $q(Q_{j})$ is from Wishart distribution $q(Q_{j})=W\left(Q_{j}\mid\hat{w}_1,\hat{\Psi}_1\right)$ in which the mean vector and covariance matrix is given by (\ref{eq:28})-(\ref{eq:29}).

\subsection{Estimation of the distribution $q\left(\alpha\right)$}
\label{sec:18} 
Similar to the previous sections, we have:

\begin{multline}\label{eq:103}
    P\left(\alpha\mid x_{1:T},z_{1:L},\pi_{1:K},\delta_{1:K},S_{1:K},Q_{1:K},\mu_{1:K},r_{1:K},A^{1:K},C^{1:K},\Xi\right)\\
    =\prod_{j=1}^{K-1}Beta\left(\pi_{j}\mid 1,\alpha\right)Gam(\alpha\mid \eta_1,\eta_2).
\end{multline}

According to the equations (\ref{eq:57}), (\ref{eq:67}) and what described before, the distribution of $q(\alpha)$ can be written as below:

\begin{equation}\label{eq:104}
    q(\alpha)\propto\exp{\left(-\alpha\Bigg(\eta_2-\sum_{j=1}^{K-1}E\left[\ln\left(1-\pi_j\right)\right]\Bigg)+\left(\eta_1-1\right)\ln\alpha\right)}.
\end{equation}

By simplifying the above equation, the distribution of $q(\alpha)$ is from Gamma and is obtained via (\ref{eq:36})-(\ref{eq:37}).
\subsection{Estimation of the distribution $q\left(a_n^{(j)}\right)$}
\label{sec:18} 
Similar to the previous sections, we have:
\begin{multline}\label{eq:104_1}
    P(a_n^{(j)}\mid x_{1:T},z_{1:L},\delta_{1:K},S_{1:K},Q_{1:K},\mu_{1:K},r_{1:K},a_{\backslash n}^{\backslash j},C^{1:K},\Psi)=\\\prod_{t=2}^T N\left(x_t^{(j)}\mid A^{(j)}x_{t-1}^{(j)},Q_j^{-1}\right)\prod_{n=1}^N N\left(a_n^{(j)}\mid 0,\sigma_A^{(j)}\right),
\end{multline}
therefore
\begin{equation}\label{eq:104_2}
    q\left(a_n^{(j)}\right)\propto\exp{E_{\backslash a_n^{(j)}} \left[\sum_{t=2}^T \ln{N\left(x_t^{(j)}\mid A^{(j)}x_{t-1}^{(j)},Q_j^{-1}\right)}+\sum_{n=1}^N \ln{N\left(a_n^{(j)}\mid 0,\sigma_A^{(j)}\right)}\right]}
\end{equation}

\begin{multline}\label{eq:104_3}
    q\left(a_n^{(j)}\right)\propto\exp{E_{\backslash a_n^{(j)}} [\sum_{n=1}^N a_n^{(j)^\intercal} \left(\hat{w}_1\hat{\Psi}_{1_{nn^\prime}} \sum_{t=2}^T E[x_{{t-1}_n}^{(j)^\intercal}x_{{t-1}_n}^{(j)}]+\sigma_A^{(j)}q(n,n) \right)a_n^{(j)} } \\
     - \sum_{n=1}^N a_n^{(j)^\intercal} \left( \sum_{n^\prime}^N \hat{w}_1 \hat{\Psi}_{1_{nn^\prime}} \sum_{t=2}^T E\left[x_{{t-1}_n}^{(j)^\intercal}x_{{t}_n^\intercal}^{(j)}\right]a_{n^{\prime}}^{(j)}     +\sum_{n^\prime}^N \hat{w}_1 \hat{\Psi}_{1_{nn^\prime}} \sum_{t=2}^T E\left[x_{{t-1}_n}^{(j)^\intercal}x_{{t}_n^\intercal}^{(j)}\right]a_{n^\prime}^{(j)}+\sum_{n^\prime=1}^N \sigma_A^{(j)}q(n,n^\prime) a_{n^\prime}^{(j)}
    \right) ]
\end{multline}

According to the equation (\ref{eq:8_1}) and expanding equation (\ref{eq:104_3}), the equations of (\ref{eq:51_1}) to (\ref{eq:51_3}) are derived.

\subsection{Estimation of the distribution $q\left(C_i^{(j)}\right)$}
\label{sec:18} 
Similar to the previous sections, we have:

\begin{multline}\label{eq:104_4}
    P\left(C_i^{(j)}\mid x_{1:T},Z_{1:L},\pi_{1:K},\delta_j,S_j,Q_{1:K},\mu_{1:K},r_{1:K},a^{1:K},C^{\backslash j}\right)=\\
    \prod_{t=1}^T N\left(y_{it}\mid C_i^{(j)}x_t^{(j)}+\mu_j,r_j{-1}\right)^{\mathbb{I}\left(z_i=j\right)}N\left(C_i^{(j)}\mid 0,\Sigma_C^{(j)}\right)
\end{multline}

therefore,
\begin{multline}\label{eq:104_4}
    q(C_i^{(j)})\propto \sum_{t=1}^T q(z_i=j)\left( C_i^{(j)} E\left[x_t^{(j)}x_t^{(j)^\intercal}\right] \hat{w}_2 \hat{\Psi}_2 C_i^{(j)^\intercal}-C_i^{(j)} E\left[x_t^{(j)}\right] \left(\hat{w}_2 \hat{\Psi}_2 y_{it} -\right.\right.\\\left.\left. \hat{w}_2 \hat{\Psi}_2\hat{m}_2\right)-\left(y_{it}^\intercal \hat{w}_2\hat{\Psi}_2 - \hat{m}_2^\intercal \hat{w}_2 \hat{\Psi}_2\right) E\left[x_t^{(j)^\intercal}\right] C_i^{(j)^\intercal} \right)-\frac{1}{2}\left(C_i^{(j)}\Sigma_{C_i}^{(j)}C_i^{(j^\intercal)}\right)
\end{multline}

According to the equation (\ref{eq:8_1}) and expanding equation (\ref{eq:104_4}), the equations of (\ref{eq:51_4}) to (\ref{eq:51_6}) are derived.

\section{Inference of the Rauch-Tung-Striebel smoother (RTSS)}
\label{sec:appB}
As mentioned before, RTSS algorithm contains two steps: Forward recursion and Backward recursion. In the Forward step, forward messages go along the LDSs.  The forward messages are defined as $\alpha_t\left(x_t^{(j)}\right)\equiv p\left(x_t^{(j)}\mid\bar{y}_{1:t}^{(j)}\right)$ , where can be written:
\begin{equation}\label{eq:105}
    \alpha_t\left(x_t^{(j)}\right)=\frac{\int{ p\left(x_{t-1}^{(j)}\mid \bar{y}_{1:t-1}^{(j)}\right) p\left(x_{t}^{(j)}\mid x_{t-1}^{(j)}\right) p\left(\bar{y}_{t}^{(j)}\mid x_{t}^{(j)}\right)}dx_{t-1}^{(j)}}{p\left(\bar{y}_{t}^{(j)}\mid \bar{y}_{1:t-1}^{(j)}\right)},
\end{equation}

where $\bar{y}_t^{(j)}$ is defined by equation (\ref{eq:49}). Suppose $\Phi_t(\bar{y}_t^{(j)})\equiv p\left(\bar{y}_t^{(j)}\mid \bar{y}_{1:t-1}^{(j)}\right)$, where

\begin{multline}\label{eq:106}
    \alpha_t(x_t)=\frac{1}{\Phi_t\left(\bar{y}_t^{(j)}\right)}\int{\alpha_{t-1}\left(x_{t-1}^{(j)}\right)p\left(x_t^{(j)}\mid x_{t-1}^{(j)} p\left(\bar{y}_t^{(j)}\mid x_t^{(j)}\right)\right)dx_{t-1}}\\
    =\frac{1}{\Phi(\bar{y}_t^{(j)})}\int{\begin{pmatrix}N\left(x_{t-1}^{(j)}\mid \dot{\mu}_{t-1},\dot{\Sigma}_{t-1}^{-1}\right)N\left(x_t^{(j)}\mid A^{(j)}x_{t-1}^{(j)},Q_j^{-1}\right)\\
    N\left(\bar{y}_t^{(j)}\mid C^{(j)}x_t^{(j)},R_j^{-1}\right)N\left(\bar{y}_t^{(j)}\mid\mu_j,\Sigma_j^{-1}\right)\end{pmatrix}dx_{t-1}^{(j)}}.    
\end{multline}

The equation (\ref{eq:70}) contains a Gaussian distribution $N\left(x_{t-1}^{(j)}\mid\ddot{\mu}_{t-1},\ddot{\Sigma}_{t-1}^{-1}\right)$, where
\begin{equation}\label{eq:107}
    \begin{array}{cc}
        & \ddot{\mu}_{t-1}=\ddot{\Sigma}_{t-1}^{-1}\left(\dot{\Sigma}_{t-1}\dot{\mu}_{t-1}+A^{(j)^\intercal}Q_jx_t^{(j)}\right)\\
        &\ddot{\Sigma}_{t-1}=\left(\dot{\Sigma}_{t-1}+A^{(j)^\intercal}Q_jA^{(j)}\right).
    \end{array}
\end{equation}
So, with respect to the above distribution, the equation (\ref{eq:70}) can be written as 

\begin{multline}\label{eq:108}
    \alpha_t(x_t)\propto N\left(\bar{y}_t^{(j)}\mid\mu_j,\Sigma_j^{-1}\right)\times\\
    \int{\exp{\frac{-1}{2}\begin{pmatrix}x_t^{(j)^\intercal}\left(Q_j+C^{(j)^\intercal}R_j C^{(j)}\right)x_t^{(j)}\\
    -x_t^{(j)^\intercal}C^{(j)^\intercal}R_j\bar{y}_t^{(j)}\\
    -\bar{y}_t^{(j)^\intercal}R_j C^{(j)} x_t^{(j)}+\dot{\mu}_{t-1}^\intercal\dot{\Sigma}_{t-1}\dot{\mu}_{t-1}\\
    +\bar{y}_t^{{(j)}^{\intercal}}R_j\bar{y}_t^{(j)}-\ddot{\mu}_{t-1}^{\intercal}\ddot{\Sigma}_{t-1}\ddot{\mu}_{t-1}\\
    \end{pmatrix}}} N\left(x_{t-1}^{(j)}\mid\ddot{\mu}_{t-1},\ddot{\Sigma}_{t-1}^{-1}
    \right) dx_{t-1}^{(j)}
\end{multline}

The distribution of the forward messages can be obtained from Normal distribution by integration of equation (\ref{eq:72}) as below:

\begin{equation}\label{eq:109}
    \begin{array}{cc}
         & \alpha_t(x_t)=N\left(x_t^{(j)}\mid\dot{\mu}_t,\dot{\Sigma}_t\right) \\
        & \dot{\mu}_t=\dot{\Sigma}_t^{-1}\left(C^{(j)^\intercal}E[R_j]\bar{y}_t^{(j)}+E[Q_j]^\intercal A^{(j)}\ddot{\Sigma}_{t-1}^{-\intercal}\dot{\Sigma}_{t-1}\dot{\mu}_{t-1}\right) \\
        & \dot{\Sigma}_t=\left(Q_j+C^{(j)^\intercal}R_jC^{(j)}-Q_j^\intercal A^{(j)}\ddot{\Sigma}_{t-1}^{-\intercal}A^{(j)^\intercal}Q_j\right).
    \end{array}
\end{equation}

In Backward step of RTSS, backward messages go back along the LDSs. The backward messages are computed as $\beta_t(x_t^{(j)})=p\left(\bar{y}_{t+1:T}^{(j)}\mid x_t^{(j)}\right)$, where can be written as 

\begin{equation}\label{eq:110}
    \beta_{t-1}(x_t^{(j)})=\int{p\left(x_t\mid x_{t-1}\right) p\left(\bar{y}_t^{(j)}\mid x_t\right)\beta_t\left(x_t^{(j)}\right)dx_t^{(j)}},
\end{equation}

in which $\beta_T(x_T^{(j)})=1$. The equation (\ref{eq:74}) can be written:
\begin{equation}\label{eq:111}
    \beta_{t-1}\left(x_{t-1}^{(j)}\right)=\int{\begin{pmatrix} N\left(x_t^{(j)}\mid A^{(j)}x_{t-1}^{(j)},Q_j^{-1}\right) N\left(\bar{y}_t^{(j)}\mid C^{(j)}x_t^{(j)},R_j^{-1}\right)\times\\
     N\left(\bar{y}_t^{(j)}\mid \mu_j,\Sigma_j^{-1}\right) N\left(x_t^{(j)}\mid\eta_t,\psi_t^{-1}\right)\end{pmatrix}dx_t^{(j)}}
\end{equation}
where it contains a Gaussian distribution $N\left(x_t^{(j)}\mid\eta_t^{\prime},\psi_t^{\prime^{-1}}\right)$, which

\begin{equation}\label{eq:112}
    \begin{array}{cc}
        & \eta_t^{\prime}=\psi_t^{\prime^{-1}}\left(Q_j A^{(j)} x_{t-1}^{(j)} + \psi_t\eta_t+ C^{(j)^\intercal} R_j \bar{y}_t^{(j)}\right) \\
        & \psi_t^\prime=\left(Q_j+\psi_t+C^{(j)^\intercal}R_j C^{(j)}\right).
    \end{array}
\end{equation}
So we can write equation (\ref{eq:76}) as:
\begin{equation}\label{eq:113}
   \exp\begin{pmatrix}\frac{-1}{2}\begin{pmatrix} x_{t-1}^{(j)^\intercal}\left(A^{(j)^\intercal}Q_j A^{(j)} -A^{(j)^\intercal} Q_j^{\intercal} \psi_t^{\prime^{-\intercal}} Q_j A^{(j)}\right) x_{t-1}^{(j)}\\
   -\left(\left(\eta_t^\intercal \psi_t^\intercal + \bar{y}_t^{(j)\intercal} R_j^\intercal C^{(j)}\right) \psi_t^{\prime^{-\intercal}} Q_j A^{(j)}\right)x_{t-1}^{(j)} \\
   -x_{t-1}^{(j)^\intercal} \left(A^{(j)^\intercal} Q_j^{\intercal} \psi_t^{\prime^{-\intercal}} \left(\psi_t \eta_t + C^{(j)^\intercal} R_j \bar{y}_t^{(j)}\right)\right) \\
   +\eta_t^\intercal \psi_t \eta_t + \bar{y}_t^{(j)^\intercal} R_j \bar{y}_t^{(j)}\\
   -\left(\eta_t^\intercal \psi_t^\intercal+ \bar{y}_t^{(j)} R_j^\intercal C^{(j)}\right)\psi_t^{\prime^{-\intercal}} \psi_t\eta_t\\
   -\left(\eta_t^\intercal \psi_t^\intercal + \bar{y}_t^{(j)^\intercal} R_j^\intercal C^{(j)}\right)\psi_t^{\prime^{-\intercal}} C^{(j)^\intercal} R_j \bar{y}_t^{(j)} \\
   \end{pmatrix}\end{pmatrix}
\end{equation}

By simplifying the equation (\ref{eq:78}), the distribution of backward messages is from  $N\left(x_{t-1}^{(j)}\mid \eta_{t-1},\psi_{t-1}^{-1}\right)$, where

\begin{equation}\label{eq:114}
    \begin{array}{cc}
        & \eta_{t-1}=\psi_{t-1}^{-1} A^{(j)^\intercal} Q_j^{\intercal} \psi_{t}^{\prime^{-\intercal}} \left(\psi_t \eta_t + C^\intercal R_j \bar{y}_t^{(j)}\right) \\
        & \psi_{t-1}=\left(A^{(j)^\intercal} Q_j A^{(j)} - A^{(j)^\intercal} Q_j^\intercal \psi_t^{\prime^{-\intercal}} Q_j A^{(j)}\right)
    \end{array}
\end{equation}

In order to compute the $1^{st}$ order and $2^{nd}$ order expected values of the hidden states we can write:

\begin{equation}\label{eq:115}
    \begin{array}{cc}
        & p\left(x_t^{(j)}\mid\bar{y}_{1:T}^{(j)}\right)\propto p\left(x_t^{(j)}\mid\bar{y}_{1:t}^{(j)}\right) p\left(\bar{y}_{t+1:T}^{(j)}\mid x_t^{(j)}\right) \\
        & =\alpha_t\left(x_t^{(j)}\right) \beta_t\left(x_t^{(j)}\right) \\
        & =N\left(x_t^{(j)}\mid\dot{\mu}_t,\dot{\Sigma}_t^{-1}\right) N\left(x_t^{(j)}\mid \eta_t,\psi_t^{-1}\right)
    \end{array}
\end{equation}

Simplifying the equation (\ref{eq:81}) yields:
\begin{equation}\label{eq:116}
    \begin{array}{cc}
        & p\left(x_t^{(j)}\mid\bar{y}_{1:T}^{(j)}=N\left(x_t^{(j)}\mid\omega_t\Gamma_{tt}^{-1}\right)\right) \\
        & \Gamma_{tt}=\dot{\Sigma}_t+\psi_t\\
        & \omega_t=\Gamma_{tt}^{-1}\left(\dot{\Sigma}_t\dot{\mu}_t+ \psi_t\eta_t\right)
    \end{array}
\end{equation}

In order to compute the $1^{st}$ order and $2^{nd}$ order expected values of the hidden states, we should compute the joint probability of $p\left(x_t^{(j)},x_{t+1}^{(j)}\mid\bar{y}_{1:T}^{(j)}\right)$. For this goal we will have:

\begin{equation}\label{eq:117}
   p\left(x_t^{(j)},x_{t+1}^{(j)}\mid\bar{y}_{1:T}^{(j)}\right)= \alpha_t(x_t^{(j)}) p\left(x_{t+1}^{(j)}\mid x_t\right) p\left(\bar{y}_{t+1}^{(j)}\mid x_{t+1}\right) \beta_{t+1}\left(x_{t+1}^{(j)}\right)
\end{equation}

Simplifying the equation (\ref{eq:82}) yields:

\begin{equation}\label{eq:118}
    p\left(x_t^{(j)},x_{t+1}^{(j)}\mid\bar{y}_{1:T}^{(j)}\right)= N
    \begin{bmatrix}
    \begin{pmatrix}
        x_t^{(j)}\\
        x_{t+1}^{(j)}
    \end{pmatrix}\!\!\big|&
    \begin{pmatrix}
        \omega_t\\
        \omega_{t+1}
    \end{pmatrix}\!\!,&
    \begin{pmatrix}
        \Gamma_{t,t} & \Gamma_{t+1,t}\\
        \Gamma_{t,t+1}^\intercal & \Gamma_{t+1,t+1}\\
    \end{pmatrix}
    \end{bmatrix}
\end{equation}

where

\begin{equation}\label{eq:119}
    \begin{array}{cc}
        & \Gamma_{t,t}=\left(\dot{\Sigma}_t+A^{(j)^\intercal} Q_j A^{(j)} + \psi_t\right)\\
        & \Gamma_{t+1,t+1}=\left(Q_j+C^{(j)^\intercal} R_j C^{(j)}\right)\\
        & \Gamma_{t+1,t}=\left(Q_j A^{(j)}\right)\\
        & \Gamma_{t,t+1}=\left(A^{(j)^\intercal} Q_j\right)
    \end{array}
\end{equation}

According to the equation (\ref{eq:83}), the $1^{st}$ order and $2^{nd}$ order expected values of the hidden states are computed as equations (\ref{eq:81})-(\ref{eq:83}).

\end{appendices}

\section{VBE}
\label{sec:appC}
\begin{equation}
    E\left[\mu_j^\intercal r_j\mu_j\right]=\frac{1}{\hat{\lambda}_2}+\hat{m}_2^2\hat{w}_2\hat{\Psi}_2
\end{equation}{}
\begin{equation}
    E\left[C_i^{(j)^\intercal}r_j C_i^{(j)}\right]=\Sigma_{C_i}^{(j)}\hat{w}_2\hat{\Psi}_2
\end{equation}{}
\begin{equation}
    E\left[\delta_j^\intercal S_j\delta_j\right]=\sum_{n^\prime=1}^N\sum_{n=1}^N \left(\frac{1}{\hat{\lambda}_3}+\hat{m}_{3_{n}}\hat{m}_{3_{n^\prime}}\hat{w}_3\hat{\Psi}_{3_{nn^\prime}}\right)
\end{equation}{}
\begin{equation}
E\left[C_i^{(j)} x_t^{(j)} x_t^{(j)^\intercal} C_i^{(j)^\intercal}\right]=\sum_{n=1}^N\sum_{n^{\prime}=1}^N \left(\hat{\Sigma}_{C_{i_{n n^\prime}}}^{(j)}+ \mu_{C_{i_{n}}}^{(j)}  \mu_{C_{i_{n^\prime}}}^{(j)}  \right) E\left[x_{t_{n^\prime}}^{(j)}x_{t_{n}}^{(j)}\right]
\end{equation}{}
\begin{multline}
E\left[ \left(x_t^{(j)}-A{(j)}x_{t-1}^{(j)}\right)\left(x_t^{(j)}-A{(j)}x_{t-1}^{(j)}\right)^\intercal\right]=E\left[x_t^{(j)}x_t^{(j)^\intercal}\right]-\mu_A^{(j)} E\left[x_{t-1}^{(j)}x_{t-1}^{(j)^\intercal}\right]-\\E\left[x_t^{(j)}x_{t-1}^{(j)^\intercal}\right]\mu_A^{(j)^\intercal}+\left\{a_n^{(j)}x_{{t-1}_n^\prime}^{(j)}a_{n^\prime}^{(j)}\right\}_{n n^\prime}
\end{multline}{}
where $\{.\}_{n n^\prime}$ indicates the $n n^\prime$-th element of matrix. 
\begin{equation}
E\left[\left(x_t^{(j)}-A^{(j)}x_{t-1}^{(j)}\right)^\intercal Q_j \left(x_t^{(j)}-A^{(j)}x_{t-1}^{(j)}\right)\right]=N^2
\end{equation}{}
\begin{equation}
E\left[x_t^{(j)^\intercal}C_i^{(j)^\intercal}r_j C_i^{(j)} x_t^{(j)}\right]=\hat{w}_2 \hat{\Psi}_2 \sum_{n=1}^N \sum_{n^{\prime}=1}^N \Sigma_{C_{i_{n n^\prime}}}^{(j)} E\left[x_{{t-1}_n^\prime}^{(j)} x_{{t-1}_n}^{(j)}\right]
\end{equation}{}
\begin{multline}
E\left[\left(y_{it}-C_i^{(j)}x_t^{(j)}-\mu_j\right)^\intercal r_j \left(y_{it}-C_i^{(j)}x_t^{(j)}-\mu_j\right)\right]=\\y_{it}^\intercal \hat{w}_2 \hat{\Psi}_2 y_{it} -E\left[x_t^{(j)^\intercal}\right]\hat{\mu}_{C_i}^{(j)^\intercal}\hat{w}_2\hat{\Psi}_2 y_{it}-\hat{m}_2^\intercal \hat{w}_2 \hat{\Psi}_2 y_{it} -y_{it}^\intercal \hat{w}_2 \hat{\Psi}_2 \hat{\mu}_{C_i}^{(j)}E\left[x_t^{(j)}\right]+\\ \hat{w}_2 \hat{\Psi}_2 \sum_{n=1}^N \sum_{n^\prime=1}^N \Sigma_{C_{i_{n n^\prime}}}^{(j)} E\left[x_{{t-1}_{n^\prime}}^{(j)} x_{{t-1}_{n}}^{(j)} \right] +\hat{m}_2^\intercal \hat{w}_2 \hat{\Psi}_2 \hat{\mu}_{C_i}^{(j)} E\left[x_t^{(j)}\right] - y_{it}^\intercal \hat{w}_2 \hat{\Psi}_2 \hat{m}_2 +\\ E\left[x_t^{(j)^\intercal}\right] \hat{\mu}_{C_i}^{(j)} \hat{w}_2 \hat{\Psi}_2 \hat{m}_2+ \frac{1}{\hat{\lambda}_2}+\hat{m}_2^2 \hat{w}_2 \hat{\Psi}_2
\end{multline}{}
\begin{multline}
E\left[(\mu_j-m_2)^\intercal \lambda_2 r_j (\mu_j-m_2)\right]=
\frac{\lambda_2}{\hat{\lambda}_2}+\lambda_2 \hat{m}_2^2 \hat{w}_2 \hat{\Psi}_2 - m_2^\intercal \lambda_2 \hat{w}_2 \hat{\Psi}_2 \hat{m}_2-\\\lambda_2 \hat{m}_2^\intercal \hat{w}_2 \hat{\Psi}_2 m_2 + m_2^\intercal \lambda_2 \hat{w}_2 \hat{\Psi}_2 m_2\end{multline}{}
\begin{equation}
E\left[x_1^{(j)^\intercal} S_j x_1^{(j)}\right]=N^2(1+\hat{\lambda}_3^{-1})+ \hat{m}_3^2 \sum_{n^\prime=1}^N \sum_{n=1}^N \{\hat{w}_3\hat{\Psi}_3\}_{n n^\prime}
\end{equation}{}
\begin{equation}
E\left[\delta_j^\intercal S_j \delta_j\right]=N^2 +\hat{m}_3^\intercal \hat{\Psi}_3 \hat{m}_3
\end{equation}{}
\begin{equation}
E\left[\delta_j^\intercal S_j x_1^{(j)}\right]= \sum_{n^\prime=1}^N \sum_{n=1}^N E\left[{x_{1_{n^\prime}}^{(j)}}\right]\hat{m}_{3_n}\hat{w}_3\hat{\Psi}_{3_{n^\prime}}
\end{equation}{}
\begin{equation}
E\left[x_1^{(j)^\intercal}S_j\delta_j\right]=\sum_{n^\prime=1}^N \sum_{n=1}^N E\left[x_{1_n}^{(j)}\right]\hat{w}_3\hat{\Psi}_{3_n}\hat{m}_{3_{n^\prime}}
\end{equation}{}
\begin{multline}
E\left[\left(x_1^{(j)}-\delta_j\right)^\intercal S_j \left(x_1^{(j)}-\delta_j\right)\right]=N^2 (1+\hat{\lambda}_3^{-1})+\hat{m}_3^2 \sum_{n^\prime =1}^N \{\hat{w}_3 \hat{\Psi}_3\}_{n n^\prime}\\ - \sum_{n^\prime =1}^N \sum_{n=1}^N E\left[x_{1_{n^\prime}}^{(j)}\right] \hat{m}_{3_n}\hat{w}_3\hat{\Psi}_{3_{n^\prime}}-\sum_{n^\prime=1}^N\sum_{n=1}^N E\left[x_{1_n}^{(j)}\right] \hat{w}_3 \hat{\Psi}_{3_n}\hat{m}_{3_{n^\prime}}+N^2+\hat{m}_3^\intercal \hat{\Psi}_3 \hat{m}_3
\end{multline}{}

\begin{multline}
E\left[\left(\delta_j-m_3\right)^\intercal \lambda_3 S_j \left(\delta_j-m_3\right)\right]=\lambda_3 \sum_{n^\prime=1}^N \sum_{n=1}^N \left(\frac{1}{\hat{\lambda}_3}+\hat{m}_{3_n} \hat{m}_{3_{n^\prime}} \hat{w}_3 \hat{\Psi}_{3_{n n^\prime}}\right)\\- m_3^\intercal\lambda_3 \hat{w}_3 \hat{\Psi}_3 \hat{m}_3 - \hat{m}_3^\intercal \hat{w}_3 \hat{\Psi}_3 \lambda_3 m_3+m_3^\intercal \lambda_3 \hat{w}_3 \hat{\Psi}_3 m_3
\end{multline}{}
\begin{equation}
E\left[a_n^{(j)^\intercal}\sigma_A^{(j)}a_n^{(j)}\right]=\sigma_A^{(j)}\left(\hat{\sigma}_{A_n}^{(j)}+\hat{\mu}_{A_n}^{(j)^2}\right)
\end{equation}{}
\begin{equation}
E\left[C_i^{(j)}\Sigma_C^{(j)}C_i^{(j)^\intercal}\right]=\sum_{n=1}^N\sum_{n^\prime=1}^N \{{\Sigma}_{C}^{(j)}\}_{n n^\prime} \left(\{\Sigma_{C_i}^{(j)}\}_{n n^\prime}-\{\hat{\mu}_{C_{i}}^{(j)}\}_{ n}\{\hat{\mu}_{C_{i}}^{(j)}\}_{ n^\prime}\right)
\end{equation}{}

\begin{multline}
E\left[\left(x_1^{(j)}-\mu_{x_1}^{(j)}\right)^\intercal\Sigma_{x_1}^{(j)}\left(x_1^{(j)}-\mu_{x_1}^{(j)}\right)\right]=\sum_{n=1}^N\sum_{n^\prime=1}^N(\{\Sigma_{x_1}^{(j)^2}\}_{n n^\prime}-\{\Sigma_{x_{1}}^{(j)}\}_{n n^\prime}\{\mu_{x_{1}}^{(j)}\}
\end{multline}{}
\begin{multline}
E\left[\left(x_1^{(j)}-\mu_{x_1}^{(j)}\right)^\intercal \Sigma_{x_1}^{(j)} \left(x_1^{(j)}-\mu_{x_1}^{(j)}\right)\right]= \\\sum_{n=1}^N \sum_{n^\prime=1}^N \left(\{\Sigma_{n_1}^{(j)^2}\}_{n n^\prime}-\{\Sigma_{x_1}^{(j)}\}_{n n^\prime}\{\mu_{x_1}^{(j)}\}_n\{\mu_{x_1}^{(j)}\}_{ n^\prime} \right)-\\2\sum_{n=1}^N \sum_{n^\prime=1}^N \{\Sigma_{x_1}^{(j)}\}_{n n^\prime} \{\mu_{x_1}^{(j)}\}_{n} E\left[x_{1_{n^\prime}}^{(j)}\right]+\sum_{n=1}^N \sum_{n^\prime=1}^N \{\Sigma_{x_{1}}^{(j)}\}_{n n^\prime} \{\mu_{x_{1}}^{(j)}\}_n\{\mu_{x_{1}}^{(j)}\}_{n^\prime}
\end{multline}{}

\begin{multline}
E\left[\left(x_t^{(j)}-\mu_{x_t}^{(j)}\right)^\intercal\Sigma_{x_t}^{(j)}\left(x_t^{(j)}-\mu_{x_t}^{(j)}\right)\right]=\sum_{n=1}^N \sum_{n^\prime=1}^N \Sigma_{x_{t_{n n^\prime}}}^{(j)} E\left[x_{t_n}^{(j)}x_{t_{n^\prime}}^{(j)}\right]\\-2\sum_{n=1}^N \sum_{n^\prime=1}^N \mu_{x_{t_n}}^{(j)} \Sigma_{x_{t_{n n^\prime}}} E\left[x_{t_{n^\prime}}^{(j)}\right]+\sum_{n=1}^N \sum_{n^\prime=1}^N \mu_{x_{t_n}}^{(j)} \Sigma_{x_{t_{n n^\prime}}}\mu_{x_{t_{n^\prime}}}^{(j)} 
\end{multline}{}

\begin{multline}
E\left[\left(\{{A^{(j)}}\}_n-\{\hat{\mu}_A^{(j)}\}_n\right)^\intercal \{\hat{\sigma}_A^{(j)}\}_n\left(\{{A^{(j)}}\}_n-\{\hat{\mu}_A^{(j)}\}_n\right)\right]=\\\{\hat{\sigma}_A^{(j)}\}_n \left(\{\hat{\sigma}_A^{(j)}\}_n-\{\hat{\mu}_A^{(j)^2}\}_n\right)-\{\hat{\mu}_A^{(j)}\}_n \{\hat{\sigma}_A^{(j)}\}_n \{\hat{\mu}_A^{(j)}\}_n
\end{multline}{}

\begin{multline}
E\left[\left(C_i^{(j)^\intercal}-\hat{\mu}_{C_i}^{(j)}\right)^\intercal \hat{\Sigma}_{C_i}^{(j)}\left(C_i^{(j)^\intercal}-\hat{\mu}_{C_i}^{(j)}\right)\right]=\\\sum_{n=1}^N \sum_{n^\prime=1}^N \{\hat{\Sigma}_{C_i}^{(j)}\}_{n n^\prime} \left(\{\hat{\Sigma}_{C_i}^{(j)}\}_{n n^\prime} - \{\hat{\mu}_{C_i}^{(j)}\}_n \{\hat{\mu}_{C_i}^{(j)}\}_{n^\prime} \right)-\sum_{n=1}^N \sum_{n^\prime=1}^N  \{\hat{\mu}_{C_i}^{(j)}\}_{n} \{\hat{\Sigma}_{C_i}^{(j)}\}_{n n^\prime}  \{\hat{\mu}_{C_i}^{(j)}\}_{n^\prime}
\end{multline}{}

\begin{equation}
E\left[\left(\mu_j-\hat{m}_2\right)^\intercal \hat{\lambda}_2 r_j \left(\mu_j-\hat{m}_2\right)\right]=1+\hat{\lambda}_2\hat{m}_2^2 \hat{w}_2\hat{\Psi}_2 -\hat{m}_2^\intercal \hat{\lambda}_2 \hat{w}_2 \hat{\Psi}_2 \hat{m}_2
\end{equation}{}




\bibliographystyle{unsrt}
\bibliography{template}

\end{document}